\documentclass[twocolumn]{emulateapj}

\usepackage{amsmath}

\shortauthors{Wong \& Sarazin}
\shorttitle{Non-Equipartition Signatures in Cluster Outskirts}
\slugcomment{Astrophysical Journal, accepted}  

\begin{document}

\title{Effects of the Non-Equipartition of Electrons and Ions in the
Outskirts of Relaxed Galaxy Clusters
}

\author{Ka-Wah Wong
and
Craig L. Sarazin
}

\affil{Department of Astronomy, University of Virginia,
P. O. Box 400325, Charlottesville, VA 22904-4325, USA
}

\email{kwwong@virginia.edu, sarazin@virginia.edu}

\begin{abstract}
We have studied the effects of electron--ion non-equipartition in
the outer
regions of relaxed clusters for a wide range of masses in the $\Lambda$CDM
cosmology using one-dimensional hydrodynamic simulations.
The effects of the non-adiabatic electron heating efficiency, $\beta$, on 
the degree of non-equipartition are also studied.
Using the gas fraction $f_{\rm gas} = 0.17$ (which is the upper limit for
a cluster), we give a conservative lower limit of the non-equipartition
effect on clusters.
We have shown that for a cluster with a mass of
$M_{\rm vir}\sim 1.2 \times 10^{15} M_{\odot}$, 
electron and ion temperatures differ
by less than a percent within the virial radius $R_{\rm vir}$.
The difference is $\approx 20\%$ for a non-adiabatic
electron heating efficiency of $\beta \sim 1/1800$ to $0.5$ at $\sim 1.4
R_{\rm vir}$.
Beyond that radius, the non-equipartition effect depends rather
strongly on $\beta$, and such a strong dependence at the shock radius can
be used to distinguish shock heating models or constrain the shock
heating efficiency of electrons.
With our simulations, we have also studied systematically the
signatures of non-equipartition on X-ray and Sunyaev--Zel'dovich (SZ) 
observables.
We have calculated the effect of
non-equipartition on the projected temperature and X-ray surface
brightness profiles using the MEKAL emission model.
We found that the
effect on the projected temperature profiles is larger than that on the
deprojected (or physical) temperature profiles.
The non-equipartition effect
can introduce a $\sim 10\%$ bias in the projected temperature at $R_{\rm
vir}$ for a wide range of $\beta$.
We also found that the
effect of non-equipartition on the projected temperature profiles can be
enhanced by increasing metallicity.
In the low-energy band $\lesssim 1$~keV, the non-equipartition model 
surface
brightness can be higher than that of the equipartition model
in the cluster outer regions.
Future X-ray observations extending to $\sim R_{\rm vir}$
or even close to the
shock radius should be able to detect these non-equipartition signatures.
For a given cluster, the difference between the SZ temperature decrements
for the equipartition and the non-equipartition models, $\delta\Delta
T_{\rm SZE}$, is larger at a higher redshift.  For the most
massive clusters at $z \approx 2$, the differences can be $\delta\Delta
T_{\rm SZE} \approx $ 4--5$~\mu$K near the shock radius.
We also found that for our model in the $\Lambda$CDM universe, the
integrated SZ bias, 
$
Y_{\rm non{\text -}eq} 
/Y_{\rm eq}$, evolves 
slightly (at a
percentage level) with redshift,
which is in contrast to the self-similar model in the Einstein--de Sitter
universe.
This may introduce biases in cosmological studies using the
$f_{\rm gas}$ technique.
We discussed briefly whether the equipartition and non-equipartition
models near the shock region can be distinguished by future radio
observations with, for example, the Atacama Large Millimeter Array.
\end{abstract}

\keywords{
cosmic microwave background ---
galaxies: clusters: general ---
hydrodynamics ---
intergalactic medium ---
shock waves ---
X-rays: galaxies: clusters
}

\section{Introduction}  
\label{sec:intro}
Observational and theoretical studies have shown that the study of the 
intracluster medium (ICM) can be used as a test of plasma physics under 
extreme environments that cannot be achieved in terrestrial laboratories, 
as well as an important cosmological probe.  If we assume the matter 
content of clusters is a fair sample of the universe, the baryon fraction 
of clusters can be used as an estimator of the average value for the 
universe, with proper correction for the baryons contained in the stellar 
component and for a small amount of baryonic matter expelled from clusters 
during the formation process.  Recent work from \citep{ARS08} has shown 
that the combined results of the baryon fraction from X-ray observation 
with the cosmic microwave background (CMB) data can give powerful 
constraints on cosmological 
parameters, such as the equation of state of the dark energy.

However, the study of cosmology using clusters of galaxies relies heavily 
on the understanding of cluster physics.  For precision cosmology, 
systematic uncertainties at even the percent level are significant.  
For example, in current studies, the baryon fraction within clusters is 
assumed to be independent of redshift and the mass of clusters.  It would 
be important to see if these assumptions are justified; if not, it is 
important to study the dependence of the baryon fraction on cluster 
properties and redshift.  Even if the dependence on redshift is weak, the 
correction factor for the baryon content within clusters compared to the 
average value in the universe could affect the constraints on cosmological 
parameters.

Studying cluster outskirts ($\gtrsim R_{200}$\footnote{
$R_{\Delta}$ is the radius within which the mean total mass density of the 
cluster is $\Delta$ times the critical density.  
The virial radius $R_{\rm vir}$ is defined as a radius within which the 
cluster is virialized.  For the Einstein--de Sitter universe, $ R_{\rm 
vir} 
\approx R_{178}$, while for the standard $\Lambda$CDM universe, $ R_{\rm 
vir} \approx R_{95}$.
}) 
is very important because the
boundary conditions of the cluster outskirts constrain the global 
properties
of a cluster \citep[e.g.,][]{TSN00}.
Also, the outer envelopes of clusters have been thought to be 
less subject to some additional physics including active galactic 
nucleus (AGN) feedback, and that 
the outer regions of clusters may provide better cosmological probes.  
Currently, there are very few observations of the properties of the ICM in 
the outer parts of clusters.  Thus, most of our understanding of these 
regions is still based on numerical hydrodynamic simulations which assume 
the hot plasma is a fluid.  In these simulations, the clusters are formed 
from mergers and accretion of dark matter and baryonic gas in overdense 
regions.  A variety of shocks with different geometries along the 
large-scale structure (LSS) filaments and transverse to them near and 
beyond the 
virial radius are unambiguous predictions of the cosmological hydrodynamic 
simulations.  Unfortunately, the lack of observational information on the 
clusters outskirts prevents us from understanding the accretion shock 
region, and hence the input physics for the numerical simulations is 
called into question.  For example, the thermodynamic state of the shocked 
gas, as well as the shock position, depends on the pre-shock gas 
temperature; the shock will be weaker if the infalling gas is pre-heated 
\citep{TSN00}.   Even worse, recently it was noted that the non-fluid 
properties may be important in regions near the virial radius, where the 
Coulomb collisional mean free path is comparable to the cluster size of a 
few Mpc \citep{Loe07}.  The Coulomb collisional timescale can also be of 
the order of the age of the cluster.  This suggests that a full kinetic 
gas theory is needed instead of the fluid approximation when studying the 
gas properties near the edge of the cluster.  Direct consequences include 
non-equipartition between electrons and ions \citep{FL97, EF98}, element 
sedimentation \citep{CL04}, and suprathermal evaporation of hot gas from 
the clusters (\citealt{Loe07}; see also \citealt{Med07}).  Some of these 
effects can lead to a bias in baryon fraction measurements, and hence 
cosmological studies.

Recently, progress has been made in the study the baryon content of the 
outer regions of clusters through the X-ray observations together with the 
Sunyaev--Zel'dovich (SZ) effect on the CMB by the 
hot electrons in the ICM out to $\sim R_{200}$ \citep{ALN+07}.  A 
3$\sigma$ result from the {\it WMAP} three-year data suggests that 
$35\%\pm8\%$ of the 
thermal energy in ICM are missing, indicating that the baryons in clusters 
may be missing even accounting for those locked in stars.  The result is 
also supported by independent measurements from other X-ray and SZ 
observations \citep{Ett03, LBC+06, VKF+06, Evr+08}.  Using the X-ray 
observations together with numerical simulations, \citet{Evr+08} reported 
that as much as $50\%$ of the thermal energy can be missing in the ICM.  
Although \citet{Gio+09} reported that the total baryon fraction within a 
smaller radius of $R_{500}$ of massive clusters are consistent to the 
cosmic value within $1 \sigma$ when all the X-ray hot gas, stellar mass in 
galaxies, gas depletion during cluster formation, and intracluster light 
from stars are taken into account, if the missing baryons measured in the 
outer region ($\lesssim R_{200}$) is really significant, this may indicate 
either a yet-unknown baryonic component,
or 
some new astrophysical processes in the ICM which is driving out the gas 
from the clusters.  While there is no evidence for any undetected baryonic 
component, \citet{ALN+07} pointed out that the missing of hot baryons can 
either be explained by the thermal diffusion or the evaporation of baryons 
out 
of the virial radius of clusters.  Another possibility is that electron 
temperature is lower than that of the equipartition value \citep{WSL+08}.

Given the advancements in the X-ray and SZ observations of the cluster 
outer 
regions \citep{ALN+07, Bau+09, GFS+09, Rei+09}, as well as the growing 
evidence of missing thermal energy in the ICM and the possible negative 
implications for cosmological tests, a more detailed study of the kinetic 
processes in cluster envelopes is necessary.  While magnetic fields may 
affect some of the kinetic effects of transport processes such as 
thermal conduction, the magnetic effects on non-equipartition should not 
be 
important since the physics is local.  Moreover, it is known that various 
astrophysical shocks in magnetized environment lead to  
non-equipartition 
\citep{GLR07, HSL+01}.

The collisionless accretion shock at the outer boundary of a cluster 
should primarily heat the ions since they carry most of the kinetic energy 
of the infalling gas.  Assuming that cluster accretion shocks are similar 
to those in supernova remnants, the electron temperature, $T_e$, 
immediately behind the shock would be lower than the ion temperature, 
$T_i$.  The equilibration between electrons and ions would then proceed by 
Coulomb collisions.  Near the virial radius, due to the low density, the 
Coulomb collisional timescale can be comparable to the age of the 
cluster, and the electrons and ions may not achieve equipartition in these 
regions \citep{FL97}.  Since X-ray and SZ observations measure the 
properties of the electrons in the ICM, the net effect is to underestimate 
the total thermal energy content within clusters.  This might account for 
some or all of the missing thermal energy in the ICM derived by the X-ray 
and SZ observations.  As mentioned above, non-equipartition of ions and 
electrons is observed in various astrophysical shocks.  Most supernova 
remnants with high Mach numbers comparable to cluster accretion shocks 
have electron temperatures which are lower than the ion temperatures 
\citep{GLR07}; in situ measurements from satellites show the same feature 
in the Earth's bow shock \citep{HSL+01}.  On the other hand, X-ray 
observations of the merger shock in the Bullet Cluster indicate that the 
equilibration time may be shorter than that expected from Coulomb 
collisions alone \citep{MV07}.  However, this merger shock has a Mach 
number of a few.  From both supernova remnant measurements and a physical 
model, \citet{GLR07} have shown that electron heating efficiency within a 
shock front (usually tens of the gyroradius) is inversely proportional to 
the Mach number squared.  If the results can also be applied to the ICM, 
the 
low Mach number merger shocks would be immediately heated to $T_e/T_i \sim 
1$, while cosmological accretion shocks with much higher Mach numbers 
would only be heated to $T_e/T_i \ll 1$ by collisionless processes.  After 
the electrons and ions pass through the thin shock front, they will likely 
be equilibrated by Coulomb collisions alone \citep{BDD08, BPP08}.

The non-equipartition in cluster of galaxies has been previously studied 
by \citet{FL97} and \citet{EF98} in semianalytic models.  They have shown 
that the temperature difference can be significant in the outer one-third 
of the shock radius of a cluster.  One- or three-dimensional simulations 
for some individual clusters have also been studied \citep{CAT98, Tak99, 
RN09}.  While these simulations use different cluster or 
cosmological models, a general agreement is that the effect of 
non-equipartition is important if shock heating efficient of electrons is 
low ($\ll 1$) and the equilibration afterward is due to Coulomb collisions 
alone.  

In this paper, we study systematically the effects of non-equipartition 
on X-ray and SZ observables in outer regions of relaxed galaxy clusters, 
which is particularly important for cosmological studies.  We carry out 
one-dimensional hydrodynamic simulations with realistic 
Navarro-Frenk-White (NFW) model under 
the concordance $\Lambda$CDM cosmological background to provide a sample 
of clusters (groups) with different masses ($10^{13}$--$10^{16} 
M_{\odot}$) 
at different redshifts ($z=0$--$2$).  
Even though we are studying the kinetic non-fluid properties in the 
cluster outer regions, the hydrodynamic treatment in modeling the cluster 
dynamical properties is reasonable and is justified as follows.  
Even dynamically unimportant magnetic fields should be able to reduce
significantly
the diffusion mean free path perpendicular to the magnetic field
\citep{Sar86,BK09}.
The suppression of diffusion in a plasma depends on the topology of magnetic 
fields.  For uniform magnetic fields, only diffusion perpendicular to the 
local magnetic field is 
suppressed, and along the field, particles move freely; their mean free 
path along a field line is still determined by Coulomb collisions. On 
large scales, diffusion is suppressed in bulk only if the magnetic 
fields are random and highly tangled on small scales. To include 
anisotropic diffusion in the calculation would be difficult since the 
magnetic field structure is not known well enough.  However, there is some 
evidence from large-scale magnetohydrodynamic simulations that magnetic 
fields in galaxy clusters are chaotic with correlation and reversal length 
scales of  $\sim 50$ and $\sim 100$~kpc, respectively \citep{DBL02}.  
Hence, we simply assume that diffusion is suppressed.  We also assume that 
electrons and ions are equilibrated locally on a long Coulomb collisional 
timescale, and assume that equilibration via plasma instabilities 
\citep{SCK+05, SCK+08} does not occur except at the shocks (see 
Section~\ref{sec:e-heating}).
Previous studies show that the dynamical properties of cluster outer 
regions in one-dimensional simulations successfully reproduce those 
simulated in 
three-dimensional calculations \citep{NFW95, RK97}.  The advantages of the 
one-dimensional simulations for our problem are presented in 
Section~\ref{sec:hydro}.
We emphasize the signatures of non-equipartition on X-ray and SZ 
observations in our studies.  We also study the effect of electron shock 
heating efficiency on the degree of non-equipartition.  Thus, observations
of electron--ion equilibration may give constraints to the electron 
heating 
efficiency, and hence the electron heating mechanism.  The wider parameter 
space compared to previous work explored in this paper allows us to study 
the impact of non-equipartition effects on cosmological studies 
in a future paper.

The paper is organized as follows.  In Section~\ref{sec:hydro}, we 
describe the 
set up of our hydrodynamic models.  The detailed implementations of the 
shock heating and the Coulomb equilibration process for our simulations 
are presented in Section~\ref{sec:e-heating}.  The ability of our 
simulations 
to reproduce analytic test models relevant to our studies is discussed in 
Section~\ref{sec:test}.  We present the simulated dynamics of our 
realistic NFW 
cluster models in the standard $\Lambda$CDM cosmology in 
Section~\ref{sec:NFW-DE_dynamics}.  These cluster models are used to 
study the 
non-equipartition effects presented in the paper.  We define the X-ray and 
SZ observables for our models to be studied, and also present the results 
for these observables in Section~\ref{sec:obs}.  We discuss and conclude 
our 
work in Section~\ref{sec:conclusion}.  Unless otherwise specified, we 
assume 
the Hubble constant $H_0 = 71.9~h_{71.9}$~km~s$^{-1}$~Mpc$^{-1}$ with 
$h_{71.9}=1$,
the total matter density parameter $\Omega_{M,0} = 0.258$, 
the dark energy density parameter $\Omega_{\Lambda} = 0.742$, 
and the gas fraction $f_{\rm gas}=\Omega_b/\Omega_M=0.17$, where 
$\Omega_b$ is the baryon density parameter, for the 
realistic NFW model in the standard 
$\Lambda$CDM 
cosmology\footnote{\tiny
http://lambda.gsfc.nasa.gov/product/map/dr3/parameters\_summary.cfm}, 
and a hydrogen mass fraction $X=76\%$ for the ICM throughout the paper.

\section{Hydrodynamic Model}
\label{sec:hydro}

LSS cosmological simulations predict that clusters do 
not evolve in isolation.  During the linear phase of structure growth, 
they can be influenced by tidal forces; while during the nonlinear growth 
phase, they can grow by accreting a significant number of smaller clusters 
or merging with clusters with similar sizes.
However, here we are interested
in studying the structure of accretion shocks in clusters, which can be
observed most easily in
relaxed clusters
which have not undergone a recent major merger.
Cosmological studies using the gas fraction in clusters are restricted 
to clusters with the highest degree of dynamical relaxation to minimize 
systematic scatter in the determination of cosmological parameters 
\citep{ARS08}; thus,
relaxed clusters are of particular interest.  
Moreover, based on a set of high resolution $N$-body simulations, it has 
been found that the mass accretion history of a dark matter halo in 
general consists of two distinct phases: an early fast phase and a late 
slow phase \citep{ WBP+02, ZJM+03, ZMJ+03, LMv+07}.  The fast accretion 
phase is dominated by major mergers, while the slow accretion phase is 
dominated by smooth accretion of background materials and many minor 
mergers.  We are most interested in studying the non-equipartition of 
electrons and ions in the outer regions of clusters, where materials 
should be continuously accreting and the morphology is roughly spherical 
symmetric.  Therefore, in our models, we simply consider cluster growth by 
smooth accretion of materials from the background cosmology.  In 
particular, we assume clusters are spherically symmetric and employ 
one-dimensional hydrodynamic simulations.  It has been shown that 
one-dimensional calculations reproduce the density and temperature 
profiles of three-dimensional simulations of clusters in the outer regions 
\citep{NFW95, RK97}, where we are
most interested.
The assumption of spherical symmetry should be
sufficient for us to gain insight into the 
astrophysical effects of non-equipartition on X-ray
and SZ observations, as well as the impact on cosmological 
studies \citep{WSL+08}.
Moreover, one-dimensional simulations also allow us 
to better resolve shocked regions and to
isolate the individual physical processes we are interested in, so 
that non-fluid properties there 
can be studied in detail.
This is difficult to achieve in three-dimensional 
simulations.  Even though three-dimensional simulations have shown that 
accretion through filaments is a general feature in related 
clusters, these simulations also show that, other than the filament 
regions, material is accreted spherically and the 
morphology in the outer skirts of a clusters are roughly spherical 
symmetric \citep{MHH+09}.  
A very recent three-dimensional study has already shown that the 
signature of non-equipartition for a relaxed cluster is roughly spherical 
symmetric as well \citep[model CL104 in][]{RN09}.

\subsection{Simulation Code}
\label{sec:code}
We employed the PLUTO code \citep{MBM+07} to solve the one-dimensional 
Newtonian hydrodynamic equations in spherical coordinates for our problem.  
This code provides a multi-physics, multi-algorithm modular environment 
which allows new physics to be included and new modules to be developed 
easily.
The code does not include self-gravity or an $N$-body solver for
determining the dark matter distribution,
but these can be handled as  
force terms in the code.  We have developed a scheme to include the dark 
matter 
contribution to the gravity by evolving the NFW profile self-consistently 
with the hydrodynamic evolution of the fluid calculated in the code 
(Section~\ref{sec:NFWevol}).  
This implementation has an advantage over using the $N$-body solver, 
in that the dynamics of gas in the NFW dark matter potential can be 
investigated under controlled conditions.
We have also implemented self-gravity of the gas in the code.  Dark energy 
can also be included in the code easily as a 
force term (Section~\ref{sec:darkenergy}).  The code is built on modern 
Godunov-type shock-capturing schemes which are particular suitable for 
computation of highly supersonic astrophysical flows in the presence of 
strong discontinuities.  
Shock-capturing is needed in our 
calculations since we are interested in calculating the non-equipartition 
signatures around the shock regions.

\subsection{Boundary Conditions and Computational Domain}
\label{sec:BC}

The boundary conditions depend on the geometry and the physics of the 
problem being solved.  For a spherically symmetric geometry of the 
ICM, 
the inner boundary condition is reflective.
Because we are studying the smooth accretion of background materials in 
isolation, it is natural to use the Hubble-flow-like outer boundary 
conditions which are defined as 
\begin{equation}
\label{eq:BCv}
\left.\frac{dv_g}{dr} \right|_{\rm in} = \left.\frac{dv_g}{dr}\right|_{\rm 
out} \, ,
\end{equation}
\begin{equation}
\label{eq:BCd}
\rho_{g,\rm in} = \rho_{g,\rm out} \, ,
\end{equation}
and
\begin{equation}
\label{eq:BCp}
P_{g,\rm in} = P_{g,\rm out} \, ,
\end{equation}
where $r$ is the radius, $v_g$, $\rho_g$, and $P_g$ are the gas velocity, 
mass density, and pressure, respectively, and the subscripts ``in'' and 
``out'' denote the inner and outer quantities at the outer boundary, 
respectively.
Each simulation is set up such that there is an over density in the 
central region compared to the critical density (Section~\ref{sec:init}).  
Near 
the over dense region, materials are accreted toward the center.  To 
avoid boundary effects, the size of the computation domain should be large 
enough that materials should always be outgoing and follow the 
Hubble-flow-like outer boundary condition.
If the size of the boundary is too small, 
the dynamics of materials near the boundary would be influenced by the 
central cluster and would not follow the Hubble-flow-like outflow; in the 
extreme case, materials should be infalling rather than outgoing.  
\citet{Ber85} has shown that in an Einstein--de Sitter universe, materials 
follow the Hubble-flow-like outflow closely
outside of a few times of the turnaround 
radius (Figure 1(a) therein).  The turnaround radius is about 10~Mpc for 
a cluster with $\sim 10^{15} M_{\odot}$ at present, and hence the 
estimated 
radius at where materials follow the Hubble-flow-like outflow is about 30~Mpc 
from the cluster center.  This condition should be sufficient 
for the $\Lambda$CDM universe, since the materials in the $\Lambda$CDM 
universe should be less bounded.  To be conservative, in our calculations, 
we take the boundary radius to be at least a few times this estimated 
radius.  In particular, we take the boundary radius to be 100~Mpc in all 
models.
The outer regions in our calculations at the final time
reproduce the 
observed Hubble flow velocity and the baryon density at redshift zero 
(Figures~\ref{fig:SS_sol_unscaled} and \ref{fig:NFW_dyn_var} 
in Sections~\ref{sec:test} and \ref{sec:NFW-DE_dynamics}, respectively), 
which 
validates our 
choice of the outer boundary conditions.

The volume of each simulation was divided into 1000 spherical annuli.
The innermost zone width is set to be 0.42~kpc.
The widths of the zones increase with radius by a fixed ratio $x$,
\begin{equation}
\label{eq:zone_ratio}
dr_{i+1} = x dr_i \, ,
\end{equation}
where $dr_i$ is the width of the $i$th zone.
The grid is defined such that
\begin{equation}
\label{eq:grid}
\sum_{i=1}^{N} dr_i = L \, ,
\end{equation}
where $N=1000$ is the total number of zones, and $L = 100$~Mpc is the size 
of the computational domain.  This is called a {\it stretched grid} in 
PLUTO.
With this grid of radii, the zones near the accretion shock
have widths of less than $1\%$ of the shock radius in 
each simulation from $z= 0 - 2$.
Doubling the resolution gives essentially the same results.

\subsection{Initial Conditions for the Realistic NFW-Dark Energy 
Cluster Models}
\label{sec:init}

Since at high enough redshift the dynamics of the background universe is 
close to the Einstein--de Sitter model, the dynamics of a cluster 
should be close to a self-similar solution as well
\citep{Ber85}.
We chose a high 
initial redshift and set up the initial conditions
to be close to the self-similar solution.
In particular, the initial condition is such 
that there are two regions separated by an accretion shock.
The location of the cluster accretion shock in the baryonic material is
very close to the first caustic in the dark matter,
and the distribution and evolution of the baryonic gas and dark matter are
almost identical outside the accretion shock
\citep{Ber85, RK97}.
Within the shock radius, we assume the dark matter 
follows the NFW profile
\citep{NFW95},
\begin{equation}
\label{eq:NFW}
\rho_{\rm dm}(r) = \frac{\rho_{\rm dm, s}}{(r/r_s)(1+r/r_s)^2} \, ,
\end{equation}
where $\rho_{\rm dm, s}$ is a density scale and $r_s$ is the scale radius.
The scale radius, $r_s$, is related to the concentration parameter, $c$, 
and the virial radius, $r_{\rm vir}$, by
\begin{equation}
\label{eq:concent}
r_s = r_{\rm vir} / c \, .
\end{equation}
The virial radius, $r_{\rm vir}$, is defined by
\begin{equation}
\label{eq:Mvir}
M_{\rm vir} = \frac{4\pi}{3} \Delta_{\rm vir} \rho_c(z) r_{\rm vir}^3 \, ,
\end{equation}
where $ M_{\rm vir}$ is the virial mass, $\rho_c(z)$ is the critical 
density of the universe at redshift $z$, and the critical overdensity 
$\Delta_{\rm vir}$ is obtained from the 
solution to the top-hat spherical collapse model, and can be approximated 
by \citep{BN98}
\begin{equation}
\label{eq:Delta}
\Delta_{\rm vir} = 18\pi^2 + 82x -39x^2 \,\,\,({\rm for \,\,} \Omega_R = 
0) \,,
\end{equation}  
where $x\equiv [\Omega_{M,0} (1+z)^3 / E(z)^2]-1$, $E(z)^2\equiv 
\Omega_{M,0} (1+z)^3+\Omega_R (1+z)^2+\Omega_{\Lambda}$, and $\Omega_R 
\equiv 1/(H_0 R)^2$ with $R$ here equals the current
radius of curvature of the universe. 
We adopt a concentration parameter 
given by equation~(\ref{eq:conc}) in Section~\ref{sec:conc}, in which $c$ 
is effectively equal to 4 in all the initial models 
considered.
We also assume
that the gas in the initial models is approximately in hydrostatic 
equilibrium with
the dark matter potential.
The choice of such initial conditions is convenient because there exists 
an analytic 
solution for isothermal hot gas in hydrostatic equilibrium with the dark 
matter potential \citep{MSS98}, but not the total potential.  This is 
convenient for setting up our initial models.
In the simulation runs, the self-gravity of gas is indeed included so that 
such initial models actually deviate from hydrostatic equilibrium 
slightly.  
We have checked that the numerical solutions we are interested in are 
rather insensitive to the set up of the initial conditions, as long as 
the initial models are approximately in hydrostatic equilibrium.
Beyond the shock radius, both the dark matter and the gas 
follow the self-similar infalling solutions.
The initial time (age of the 
universe) is chosen to be
29.4~Myr, which corresponds to a 
redshift of $z = 70.6$ in
the standard $\Lambda$CDM cosmology.
The details of the setup are described below.

We first define an initial cluster mass, $M_{\rm sh,\it i}$, at the 
initial time chosen.  The initial cluster mass is distributed within an 
initial shock radius, $R_{\rm sh, \it i}$.  
In the self-similar solution \citep{Ber85}, $R_{\rm sh, \it i}$ and 
$M_{\rm sh,\it i}$ are related by
\begin{equation}
\label{eq:init_radius}
R_{\rm sh, \it i} = \lambda_{\rm scale} r_{\rm ta, \it i} \, ,
\end{equation}
and
\begin{equation}
\label{eq:init_mass}
M_{\rm sh, \it i} = \frac{4\pi}{3} \rho_{c,i} \,  r_{\rm ta, \it i}^3 
m_{\rm scale} \, , 
\end{equation}
where 
$r_{\rm ta, \it i}$ is the initial turnaround radius, 
$\rho_{c,i}$ is the critical density at the initial time, 
and,
$\lambda_{\rm scale} = 0.347$ and $m_{\rm scale} = 3.54$ are the 
dimensionless scaled 
radius and mass in the self-similar solution at the shock radius, 
respectively.
This implies that the initial average overdensity within the accretion 
shock is\begin{equation}
\label{eq:overdensity_shock}
\Delta_{\rm sh,\it i} = m_{\rm scale}/\lambda_{\rm scale}^3 = 84.73
\, .
\end{equation}
Within the 
shock radius, we assume the dark matter distribution is given by the NFW
profile (equation~\ref{eq:NFW}).
For a fixed value of concentration parameter, $c=4$ assumed at high 
redshift (Section~\ref{sec:conc}), the initial scale radius $r_{s,i}$ can 
be solved by using equation~(C9) in \citet{HK03}, which is
\begin{equation}
\label{eq:HK03eqC9}
\frac{r_{s,i}}{R_{\rm sh,\it i}} = x\left[ f_{\rm sh,\it i} = 
\frac{\Delta_{\rm 
sh, \it i}}{\Delta_{\rm vir, \it i}} f(1/c) \right],
\end{equation}
where $f(x)=x^3[\ln(1+x^{-1})-(1+x)^{-1}]$, $x(f)$ in an 
accurate fitting form is given in Appendix~C in \citet{HK03}, 
and $\Delta_{\rm vir}$ is given in equation~(\ref{eq:Delta}).
In general, $c$ depends on mass, and $r_s$ can be calculated iteratively.
The scale density $\rho_{\rm dm, s}$ 
in equation~(\ref{eq:NFW})
is then fixed by requiring that the total dark matter 
integrated to the shock radius is equal to $(1-f_{\rm gas}) M_{\rm sh, \it 
i}$.

Within the shock radius, we assume the gas is initially in 
approximately hydrostatic 
equilibrium ($v_g = 0$) with the dark matter potential.
Initially, the gas is assumed to be isothermal.
The gas distribution can be solved 
analytically, and is given by \citep{MSS98}
\begin{equation}
\label{eq:rho_g}
\rho_g(r) = \rho_{g, 0} \, e^{-27 b /2} \left( 1+\frac{r}{r_s} 
\right)^{27b/(2r/r_s)} \, .
\end{equation}
The central gas density $\rho_{g, 0}$ and $b$ are fixed by 
simultaneously requiring that 
the gas fraction within the shock radius is equal to $f_{\rm gas}$ and the 
gas density at the shock radius is given by the strong shock jump 
condition
\begin{equation}
\label{eq:rho_shock}
\rho_{g2} = 4 \rho_{g1}
\, ,
\end{equation}
where the subscripts 1 and 2 denote the preshock and postshock quantities, 
respectively.
Once $b$ is fixed, the temperature, and hence the gas pressure is indeed 
fixed by using equation~(9) in \citet{MSS98}.  The gas pressure, $P_g$, is 
proportional to $\rho_g$ since the gas is assumed to be isothermal.
However, if such pressure profile 
is used, the pressure at the shock radius does not match the strong shock 
jump condition in general.  This is because the dynamical solution with 
accretion shock does not follow the isothermal hydrostatic solution given 
in \citet{MSS98}.  Since we are interested in the shock solution in our 
calculation, we renormalize the gas pressure such that the strong shock 
jump condition is satisfied at the shock radius,
\begin{equation}
\label{eq:P_g}
P_{g2} = \frac{4}{3} \rho_{g1} \,  v_{g1}^2 \, .
\end{equation}
With this renormalization, the pressure is only lower than the 
isothermal hydrostatic equilibrium pressure by 17\%.  The initial gas 
deviates slight from hydrostatic equilibrium.  As long as the gas is in 
approximately equilibrium initially, the late time evolution of the 
cluster profiles in the outer regions we are interested in should not be 
affected by the slightly deviation in the initial condition.

Gas in exact hydrostatic equilibrium with the total gravitational 
potential satisfying both the density and pressure shock jump conditions 
is not isothermal in general.  We have checked that using initial 
conditions in exact hydrostatic equilibrium give essentially the same 
results.

Outside the shock radius, the gas density and velocity profiles are assumed
to follow
the self-similar infall solution \citep{Ber85}, and the pressure is 
set to be effectively zero (i.e., the smallest value the code allows).

\subsection{Concentration Parameters}
\label{sec:conc}

We also need to set the value of the concentration parameter $c$.
\citet{BKS+01} proposed that $c$ scales as $(1+z)^{-1}$.
At low redshift, this is supported by numerical simulations by 
\citet{WBP+02} and \citet{ZMJ+03, ZJM+09}, when the dark matter halos are 
in the slow accretion phase.  However, at high enough redshift when the 
dark matter halos are in the fast accretion phase, the concentration 
parameter approaches a constant minimum value independent of the mass 
and redshift of the halo \citep{ZJM+03, ZJM+09, GNC+08}.
At low redshifts, the variation of the concentration parameter with mass
can simply be fitted using a power law \citep{DBP+04, Net+07}.  
Thus, we adapt the concentration parameter used by \citet{BGH+07} but set 
a minimum value of 4 \citep{ZJM+09}
independent of the redshift and the halo mass of the cluster in our 
simulations:
\begin{equation}
\label{eq:conc}
c(M_{\rm vir},z) = \max \left[\frac{c_{14}}{1+z} \left(\frac{M_{\rm 
vir}}{M_{14}}\right)^{\alpha},4\right] \, ,
\end{equation}
where we take
$c_{14}=9.0$ and $\alpha=-0.172$ as determined from the X-ray galaxy 
clusters with halo masses between $(0.06$--$20) \times 10^{14} M_{\odot}$ 
at low redshifts
\citep{BGH+07} and
$M_{14} \equiv 10^{14} h^{-1}_{100} 
M_{\odot}$.

\subsection{Self-consistent Evolution of NFW Potential Including Accreted 
Mass}
\label{sec:NFWevol}

At each time step, we determine the gas shock radius, and we assume dark 
matter is distributed according to the NFW profile within the shock 
radius; outside that radius, the dynamics of dark matter and gas are 
the same.  We assume 
the baryon fraction is conserved as materials accretes within the shock,
and hence, the dark matter mass within the shock radius is equal to 
($1/f_{\rm gas}-1$) times the gas mass within that region.  Although 
three-dimensional simulations suggest that there is a hydrodynamic outflow 
of gas and $f_{\rm gas}$ inside a cluster is smaller than the cosmological 
value, our treatment here is at least self-consistent.  Moreover, a
smaller value of $f_{\rm gas}$  inside a cluster in reality would enhance the 
effect of non-equipartition, and hence our calculations give a conservative 
estimate of the non-equipartition effect.  As mentioned in 
Section~\ref{sec:init}, if the shock radius and the mass within it are 
known, 
the dark matter profile can be solved iteratively together with the 
concentration parameter.  The dark matter profile within the shock radius 
calculated in each time step is used to calculate the gravity contributed 
by the dark matter within the shock radius.

\subsection{Dark Energy Implementation}
\label{sec:darkenergy}

We will do the calculations in a cosmological background including the
accelerated expansion due to dark energy.
The accretion of materials depends sensitively on dark energy since it 
governs the expansion of the background materials which are to be accreted 
\citep{RK97}.  To implement dark energy in the one-dimensional Newtonian 
code, we utilize the similarity between the Friedmann solution and 
Newtonian cosmology, since the latter can be solved in the Newtonian 
code.
The dynamical solution of the scale factor, $a$, of a homogeneous
universe is governed by
\begin{equation}
\label{eq:scalefactor_eq}
\ddot{a} =-\frac{4\pi G}{3} a  \left(\rho + \frac{3P}{c^2} \right) \, ,
\end{equation}
where $\rho$ and $P$ are the total energy density and pressure of the 
universe, respectively, and $G$ and $c$ are the gravitational constant and 
speed of light, respectively.  The Newtonian analog is simply replacing 
$a$ by $r$, i.e.
\begin{equation}
\label{eq:newtonian_eq}
\frac{d^2 r}{d t^2} = -\frac{4\pi G}{3} r \left(\rho + \frac{3P}{c^2} 
\right) \, .
\end{equation}
For a pure dark energy universe where the dark energy is described by the 
cosmological constant $\Lambda$, $\rho = \rho_{\Lambda}$, $P = 
P_{\Lambda}$, and $\rho_{\Lambda} = -P_{\Lambda}/c^2$.  Thus, 
equation~(\ref{eq:newtonian_eq}) becomes
\begin{equation}
\label{eq:DEforce}
\frac{d^2 r}{d t^2} = \frac{8 \pi G}{3} \rho_{\Lambda} \, r \,.
\end{equation}
We identify the right-hand side of equation~(\ref{eq:DEforce}) to be the 
repulsive force per unit mass contributed by the dark energy.  The dark 
energy density, $\rho_{\Lambda}$, in the $\Lambda$CDM cosmology is 
constant throughout the history of the universe, and is given by
\begin{equation}
\label{eq:rho_L}
\rho_{\Lambda} = \Omega_{\Lambda,0} \, \rho_{c,0} \, ,
\end{equation}
where the quantities on the right-hand side are evaluated at the present 
time.  With the force term in equation~(\ref{eq:DEforce}) implemented into 
the code, the baryon density of the background cosmology is reproduced 
correctly in each of our simulations (see Figure~\ref{fig:NFW_dyn_var} 
below).

\section{Electron Heating within a Cluster}
\label{sec:e-heating}

\subsection{General Picture of Shock Heating and the Equilibration between 
Electrons and Ions Afterward}
\label{sec:e-heating_general}

The heating of electrons passing through a shock within a cluster involves 
at least a two-step process.  The first step is the shock heating of 
electrons within the very narrow shock, and the second step is the 
equilibration between electrons and ions afterward.  In this section, we 
outline the general picture of electron heating within a cluster we 
assumed for our model.  The detailed implementations of the shock heating 
(first step) and the Coulomb equilibration afterward (second step) are 
presented in Sections~\ref{sec:shock_heating} and \ref{sec:eq_coul}, 
respectively.  Section~\ref{sec:steps} outlines the steps to calculate 
electron heating in our code.

Immediately after the infalling material (electrons and ions, together 
with dark matter) has passed through a collisionless accretion shock, we 
assume it should primarily heat the ions rather than electrons since they 
carry most of the kinetic energy of the infalling gas.  
The electron 
temperature immediately behind the shock would be lower than the ion 
temperature, and they would be in a state of non-equipartition.  Indeed, 
non-equipartition of ions and electrons is known in various astrophysical 
shocks.  Most supernova remnants with high Mach numbers comparable to 
cluster accretion shocks have electron temperatures which are lower than 
the ion temperatures \citep{GLR07}; in situ measurements from satellites 
show the same feature in the Earth's bow shock \citep{HSL+01}.  It is 
reasonable to assume that cluster accretion shocks are similar to those 
astrophysical shocks.  
Within the very narrow shock region, electrons and ions can
exchange energy by collisionless processes generated by plasma 
instabilities there.  Unfortunately, the collisionless heating 
mechanisms and the rate of heating are still poorly known theoretically.  
We simply model the shock heating efficiency for the electron as a free 
parameter (Section~\ref{sec:shock_heating}).

After the electrons and ions have passed through the thin shock 
front, they 
will then be equilibrated by Coulomb collisions \citep{BDD08, BPP08}.
It is also possible that behind the shock, 
electrons and ions can
exchange energy by collisionless processes generated by plasma 
instabilities left behind.
Again, 
the collisionless heating mechanisms and the rate of equilibration 
are still poorly known theoretically.
It has been argued that behind a strong shock, the spectra of supernova 
remnants 
are consistent with purely Coulomb equilibration processes \citep{RGH03}.
Given both the theoretical uncertainty concerning the 
collisionless heating mechanisms and at least some observational evidence 
that collisionless processes are not important in the equilibration
downstream,
we simply consider the Coulomb collisions as the only 
equilibration process behind the shock in our calculations 
(Section~\ref{sec:eq_coul}).

In calculating the equilibration processes of electrons and ions, we 
consider a fully ionized plasma of hydrogen and helium only.  
The timescale for charged particles of the same species to achieve a 
Maxwellian distribution can be estimated by the collisional timescale 
\citep[p. 133]{Spi62}
\begin{widetext}
\begin{equation}
\label{eq:txx}
t_{xx}=28.6\,{\rm Myr}\,  \left(\frac{T_x}{10^7\,{\rm K}}\right)^{3/2} 
\left( \frac{n_x}{10^{-5}\, {\rm cm}^{-3}}   \right)^{-1}  \left( 
\frac{\ln \Lambda}{40} \right)^{-1} \frac{A_x^{1/2}}{Z^{4}_{x}}  \, ,
\end{equation}
\end{widetext}
where $T_x$ is the temperature of the species $x$ after it has achieved 
equilibrium, $n_x$ is the number density, $A_x$ is the particle mass 
number, $Z_x$ is the particle charge, and $\ln \Lambda$ is the Coulomb 
logarithm which is similar for all species of interest.  
For the charged particles we are interested in, the electron--electron 
collisional timescale is the shortest, with $ t_{ee} \sim t_{pp} A_e 
^{1/2} \sim t_{pp} /43$.
For typical abundance $n_{\rm H} \sim 10~n_{\rm 
He}$,
$t_{pp} \sim t_{\rm He^{+2} He^{+2}} $.  
For protons, the collisional timescale is $t_{pp} \sim 30$~Myr, which 
is much shorter than the age of the cluster or the accretion timescale.  
For an accretion shock propagating outward with a velocity $v_s \sim 
1000$~km~s$^{-1}$, protons at a distance
$d \sim v_s  t_{pp} / 4 \sim 7$~kpc 
away from the shock should be in a Maxwellian distribution.  This is of the 
order of the zone size ($\sim 10$~kpc) at the shock radius in our 
calculations.  Hence, we can assume that protons, helium ions, and 
electrons are in equilibrium independently with temperatures $T_p$, 
$T_{{\rm He}^{+2}}$, and $T_e$, respectively.  
The energy exchange timescale for two species of particles, $x$ and $y$
in Maxwellian distributions with temperatures $T_x$ and $T_y$ is given by 
the equilibration timescale \citep[p. 135]{Spi62}
\begin{widetext}
\begin{equation}
\label{eq:teq}
t_{xy}=14.7\,{\rm Myr}\, \left(\frac{T_x}{10^7\,{\rm K}}\right)^{3/2}  
\left( \frac{n_y}{10^{-5}\, {\rm cm}^{-3}}   \right)^{-1}  \left( 
\frac{\ln \Lambda}{40} \right)^{-1} \frac{A_y}{Z_{x}^{2} Z_{y}^{2} 
A_x^{1/2}} 
\left(  1+\frac{T_y}{T_x}\frac{A_x}{A_y} \right)^{3/2} \, .
\end{equation}
\end{widetext}
Consider energy exchange between electrons and protons, and assume that the
electrons are initially much cooler than the protons.
When $T_e \la T_p (m_e/m_p)$, the electrons are heated rapidly, with
$t_{ep} \sim t_{pp} / 3600$.
Thus, we expect that most of the time spent in electron heating will
occur in the regime in which $T_e/T_p \gg 
m_e/m_p$.  Hence, the energy exchange timescale between electrons and 
protons, $t_{ep}$  is on the order of $0.5 (m_p/m_e)^{1/2} t_{pp} \sim 21
t_{pp}$.
Similarly, $t_{e{\rm He}^{+2}} \gtrsim t_{ep}$ for $n_{\rm H} \sim 
10~n_{\rm 
He}$.
The equipartition timescale between electrons and ions 
(either protons or helium ions) will be of the order of $0.6$~Gyr, which 
is comparable to the accretion timescale of a cluster,
[(cluster size)/(accretion velocity)] $\sim 1$~Gyr.
Thus, $T_e$ may not be equal to 
$T_p$ or $T_{{\rm He}^{+2}}$ in the outer regions of a cluster, which is 
the problem we are considering.
From equation~(\ref{eq:teq}), it can be seen that $t_{{\rm He}^{+2}p} 
\gtrsim t_{pp}$.  It is also possible that $T_p \neq T_{{\rm He}^{+2}}$ in 
the regions of interest.  
The solutions of the equilibration of the plasma formally can be obtained 
by solving the equilibration between the three species.
If we are 
interested only in the electron temperature evolution, we can consider the 
energy exchange rate between the electrons and both of the two ion species
\begin{equation}
\label{eq:tei}
1/t_{ei} = 1/t_{ep}+1/t_{e{\rm He}^{+2}} \, .
\end{equation}
Since the contribution per ion to $t_{ei}^{-1}$ is the same for protons 
and helium ions (equation~\ref{eq:teq}), the rate of change of electron 
temperature with the plasma, $dT_e/dt$, depends only on the mean ion 
temperature \citep{FL97}.  Moreover, once the ions reach equilibrium with 
each other, the rate of change of their temperatures due to collisions 
with electrons is the same, so they will remain in equilibrium.  Thus, for 
simplicity, we assume a single ion temperature, $T_i$, and we assume the 
electrons with temperature $T_e$ are equilibrated with this single ion 
temperature plasma.

In hydrodynamic simulations, the three hydrodynamic variables, the 
gas mass density, $\rho_g$, the total gas pressure, $P_{g}$, and the gas 
velocity, $v_g$, determine the hydrodynamic state of the fluid 
completely, independent of the kinetic state of the plasma.
Since we are 
interested in the difference between electron temperature, $T_e$, and ion 
temperature, $T_i$, we assign a new variable, $\tau \equiv 
T_e/{\bar {T}}$, in each grid of our hydrodynamic calculations, where 
${\bar{T}}$ is the average thermodynamic temperature of the fluid given by 
${\bar{T}} \equiv (\mu m_p/k_B)(P_g/ \rho_g)$, 
where $\mu=0.59$ is the mean molecular weight, $m_p$ is the proton mass, 
and $k_B$ is the Boltzmann constant.
The electron temperature, 
$T_e$, the ion temperature, $T_i$, and the average thermodynamic 
temperature, ${\bar{T}}$, are simply related by
\begin{equation}
\label{eq:barT}
{\bar{T}} = \frac{n_e T_e + n_i T_i}{n_e + n_i}
\, ,
\end{equation}
where $n_e$ and $n_i$ are the electron and ion number densities, 
respectively.

\subsection{Electron Heating within the Thin Shock Front}
\label{sec:shock_heating}

Within the very narrow shock front, electrons can be heated by both 
adiabatic 
compression and non-adiabatic heating.  Non-adiabatic heating includes the 
conversion of the bulk kinetic energy into thermal energy
and other collisionless heating processes generated by plasma instabilities.
We can 
define the total non-adiabatic electron heating efficiency in the shock, 
$\beta$, to be the change in electron temperature due to non-adiabatic 
heating, $\Delta T_{e, \rm non{\text -}ad}$, relative to the change in 
average 
thermodynamic temperature due to non-adiabatic heating, $\Delta 
\bar{T}_{\rm non{\text -}ad}$,
\begin{equation}
\label{eq:beta}
\beta \equiv \left( \frac{\Delta T_e}{\Delta \bar{T}}  \right)_{\rm 
non{\text -}ad}
\, . 
\end{equation}
The change in the bulk kinetic energy per particle is
$m(v_{g1}^2-v_{g2}^2)/2$, where $m$ is the particle mass,
and hence the increase in the electron temperature is much 
smaller than that of the ions in a shock.
For the electron heating processes generated by plasma 
instabilities, the details of the mechanism and the efficiency are still 
unclear.  Therefore, in our study, we assume total non-adiabatic electron 
heating efficiency, $\beta$, to be a constant
for any given cluster model.
Adiabatic compression increases the electron temperature by a factor of 
$(\rho_{e2} / \rho_{e1})^{\gamma -1}$, where $\gamma $ is the adiabatic 
index, $\rho_{e1}$ and $\rho_{e2}$ are the preshock and postshock electron 
temperatures, respectively.
For a strong shock with $\gamma=5/3$, adiabatic compression only 
increases the electron temperature by a factor of
$2^{4/3} \approx 2.5$.
Thus, if the 
preshock electron temperature, $T_{e1}$, is negligible compared to the 
postshock average temperature, ${\bar{T_2}}$, adiabatic heating would 
not be important.  This is indeed the case in our models.  Nevertheless, we 
have included adiabatic heating in our calculations.  Including both 
adiabatic and non-adiabatic heating, the postshock electron temperature 
immediately after the shock
is given by (Appendix~\ref{app1})
\begin{equation}
\label{eq:shockheating}
T_{e2} = \left(\frac{\rho_{g2}}{\rho_{g1}}\right)^{\gamma -1} T_{e1} +
\beta\,
\max \left[ 0
,\bar{T_2}  - \left(\frac{\rho_{g2}}{\rho_{g1}}\right)^{\gamma -1}
\bar{T_1} \right]
\, ,
\end{equation}
where a minimum of zero in the second term is set to ensure that numerical 
fluctuations do not introduce a false decrease in entropy in the 
non-adiabatic heating.

Note that we assume that the heating of electrons is equivalent to the 
increase of temperature.  This is only true if the electron distribution 
is Maxwellian.  The time for the electrons to achieve a Maxwellian 
distribution after they have been heated at the shock is $t_{ee} \sim 
t_{pp} / 43 \sim 0.7 \, \rm{Myr}$ (Section~\ref{sec:e-heating_general}).  
As shown in Section~\ref{sec:e-heating_general}, this is much shorter 
than the accretion 
timescale, and the electrons within the zone size at the shock radius of 
our simulations should achieve the Maxwellian distribution.
Note also that even if electrons are in non-Maxwellian distribution and if 
such electrons are heated much faster than 
the Coulomb collisional heating rate, the faster heating within the shock has 
already been parameterized by the electron heating efficiency $\beta$.  

Recently, observations of supernova remnants have shown that the electron 
heating efficiency at the shock is inversely proportional to the Mach 
number squared 
\citep{GLR07}.  If the result can be applied to cluster accretion shocks, 
the high accretion shock Mach number (${\cal M} > 100$) would imply the 
electron heating efficiency to be $\ll 1$ in cluster accretion shocks.
In our work, we consider two cases for the shock heating of electrons:
$\beta = 1/1800$ (the mass ratio of electrons to protons) as a model
for a very low non-adiabatic heating efficiency which is supported by
supernova remnant observations, and $\beta = 0.5$ as a model for
an intermediate heating efficiency within the shock.

\subsection{Coulomb Equilibration After Shock Heating}
\label{sec:eq_coul}

After the shock-heated material has passed through the thin shock front, 
we assume Coulomb equilibration as the only heating process for electrons 
behind the shock.
The evolution of electron temperature due to Coulomb collision is given by 
\citep[p. 135]{Spi62}
\begin{equation}
\label{eq:equilibration}
\frac{d \tau}{dt}  = \frac{2 \ln \Lambda}{503} \left< \frac{Z^2}{A} 
\right> \frac{n}{\bar{T}^{3/2}} \tau^{-3/2} ( 1-\tau) \, {\rm 
s}^{-1} \, ,
\end{equation}
where $n$ is the total particle number density,  $\ln \Lambda \approx 37.8 
+ \ln (T_e/10^8 {\rm K})-\ln (n_e/10^{-3}{\rm cm}^{-3})^{1/2}$ is the 
Coulomb logarithm, the angle bracket term is the mean value over the ratio 
of the square of ion charge $Z$ and the atomic number $A$.
For our model 
of a pure hydrogen and helium gas, the angle bracket term equals 1
as long as $T_e/T_i \gtrsim m_e/m_p$, which is true for our models with 
a minimum $\beta = 1/1800$.
We solve equation~(\ref{eq:equilibration}) for each time step in our 
hydrodynamic calculations (Section~\ref{sec:steps}).

\subsection{Steps to Calculate the $T_e/\bar{T}$ Evolution}
\label{sec:steps}

Cosmological accretion shocks are identified 
as shocks in which the Mach number, ${\cal M}$, is greater than 10.
In our simulations, we use the pressure shock jump condition,
\begin{equation}
\label{eq:pshockjump}
\frac{P_{g2}}{P_{g1}} = 
\frac{2\gamma}{\gamma+1}{\cal M}^2-\frac{\gamma-1}{\gamma+1}
\, ,
\end{equation}
to identify shocks.
In each time step of our hydrodynamic simulations, if the condition of 
${\cal M} > 10$ in equation~(\ref{eq:pshockjump}) is satisfied, we 
identify there is a strong shock at $r_2$.  We found that this 
condition can identify cosmological shocks correctly in our simulations. 
In regions without a strong shock, we 
calculate the $T_e/\bar{T}$ evolution using 
equation~(\ref{eq:equilibration}) throughout the entire time step.  If an 
accretion shock is identified, we first use 
equation~(\ref{eq:equilibration}) to evolve $T_e/\bar{T}$ for one half of 
the time step.  Then, we apply the shock heating using 
equation~(\ref{eq:shockheating}).  Finally, we use 
equation~(\ref{eq:equilibration}) again to evolve $T_e/\bar{T}$ for the 
reminding half of the time step.

Cosmological simulations show that there are various kinds of shocks 
formed in the LLSs \citep{KRC+07}.   Based on their 
location, these shocks can be classified as external and internal shocks.  
External shocks are formed around the outermost surfaces of the LLSs that 
unshocked intergalactic gas is falling onto (sheets, 
filaments, and halos), and the gas is shock heated for the first time.  
These external shocks in general have Mach numbers $\gg 10$.  
Internal shocks have low Mach numbers $\lesssim 10$.  They are 
formed within those nonlinear structures, e.g., galaxy clusters and 
filaments, by the infall of previously shocked gas during subclump 
mergers, as well as by chaotic flow motions.
These cosmological simulations show that there are a large number of low 
Mach number internal shocks contributing significantly to the thermal 
energy budget of a cluster \citep{KRC+07}.
Since those small Mach number internal shocks are mainly found well within inner 
regions of a cluster where the densities are high enough, electrons and 
ions are in general in equipartition.  
Here, we are only interested in the non-equipartition effects of electrons 
and ions, and these effects are only significant in very low density 
regions (e.g., cosmological accretion shock regions) rather than the 
contribution of the thermal energy from the shocks to the cluster.
For our one-dimensional simulations which are used to study relaxed 
clusters, due to the symmetry of the problem, there is in fact only one 
cosmological accretion shock with Mach number $\gg 10$ for each cluster, 
and there are no internal shocks formed.  For comparison, 
three-dimensional 
simulations have shown that for relaxed clusters, the number of small Mach 
number internal shocks is much smaller than that of the high Mach number 
external shocks \citep{MHH+09}.  This is because internal shocks are 
formed mostly by mergers or chaotic flow motions which are less likely to 
be found in relaxed clusters by definition.  Thus, we do not have to 
consider electron heating by the low Mach number internal shocks at least 
for relaxed cluster that we are interested in.  Future three-dimensional 
simulations will be needed to test our assumption.

\section{Test Models: Self-similar Models}
\label{sec:test}
The essential physics governing the equilibration problem we are 
considering are the hydrodynamics of the gas, the gravity by both dark 
matter and gas, the equilibration physics between electrons and ions, as 
well as the dark energy which modifies the background cosmology, and hence 
the rate of accretion onto clusters and the duration of the 
accretion 
throughout cosmic history.  The contribution of dark energy can be 
tested by whether the baryon density of the background cosmology can be 
reproduced correctly in each of our simulations, and this is indeed the 
case for our realistic simulations which will be presented in 
Section~\ref{sec:NFW-DE_dynamics} (see Figure~\ref{fig:NFW_dyn_var} 
below).  We 
present three tests here to show that our hydro code can handle the 
necessary physics correctly for the problem we are considering.  The 
hydrodynamical response of the gas in an external gravitational field is 
tested by 
reproducing the analytic self-similar solutions of a collisionless dark 
matter dominated accretion model \citep{Ber85}.  The correct handling of 
self-gravity of the gas is tested by reproducing the analytic self-similar 
solutions of a collisional gas dominated accretion model \citep{Ber85}.  
Finally, the correct handling of the equilibration physics is tested by 
comparing to the analytic solutions of a self-similar non-equipartition 
model calculated by \citet{FL97}.

\subsection{Collisionless Dark Matter Dominated Accretion Model}
\label{sec:test_ex-gravity}
\begin{figure}
\includegraphics[angle=270,width=8.2cm]{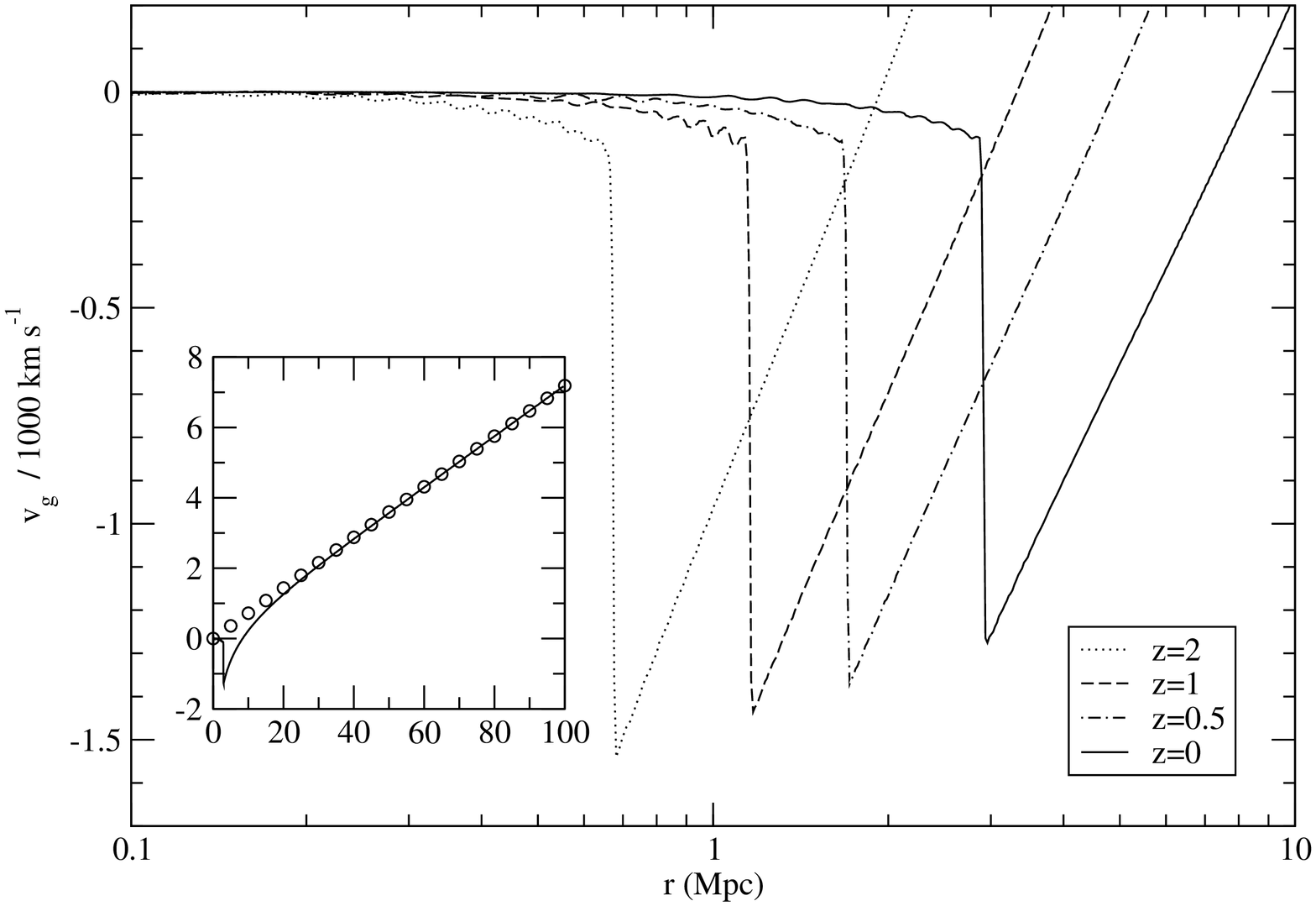}
\\
\includegraphics[angle=270,width=8.2cm]{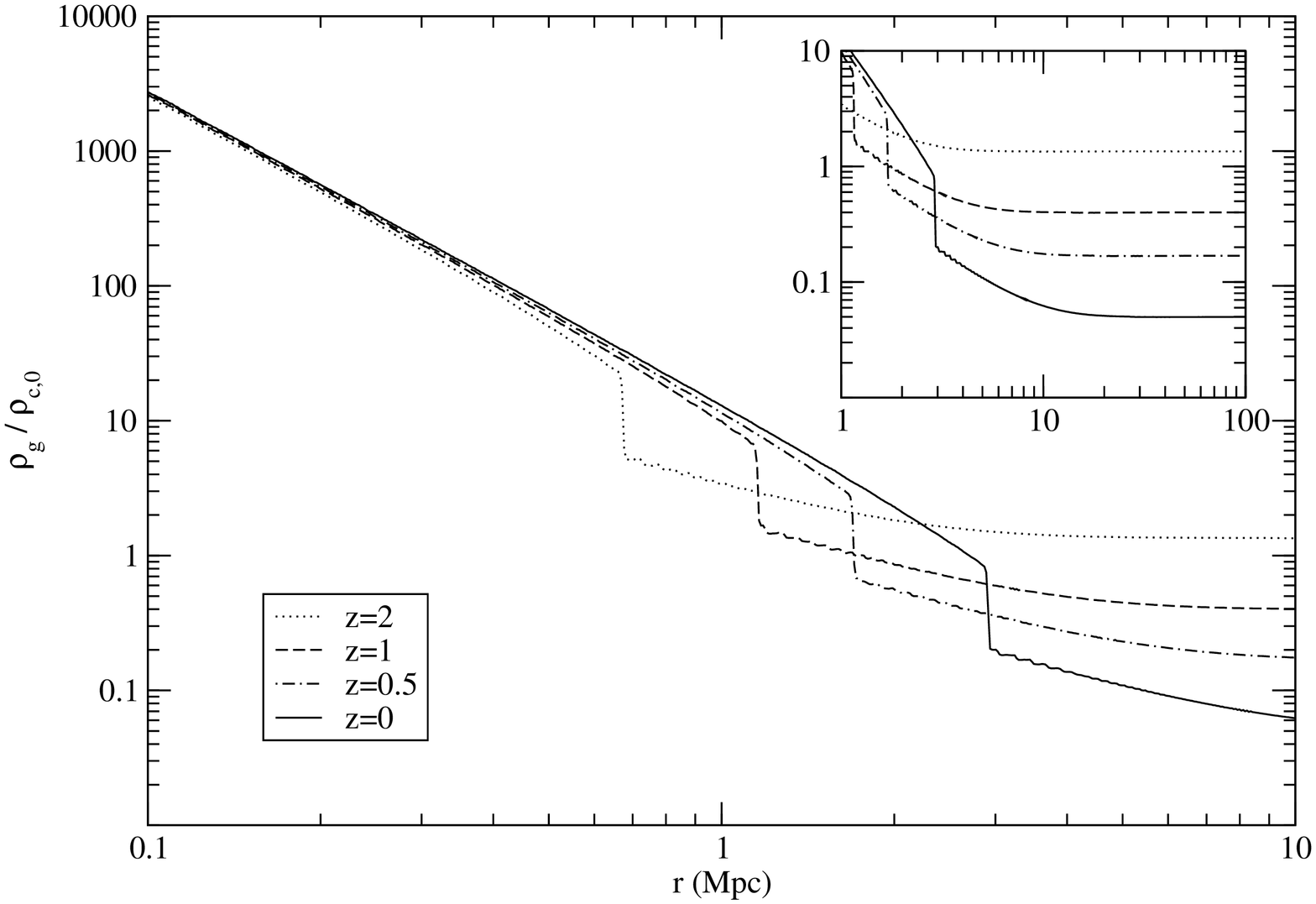}
\\
\includegraphics[angle=270,width=8.2cm]{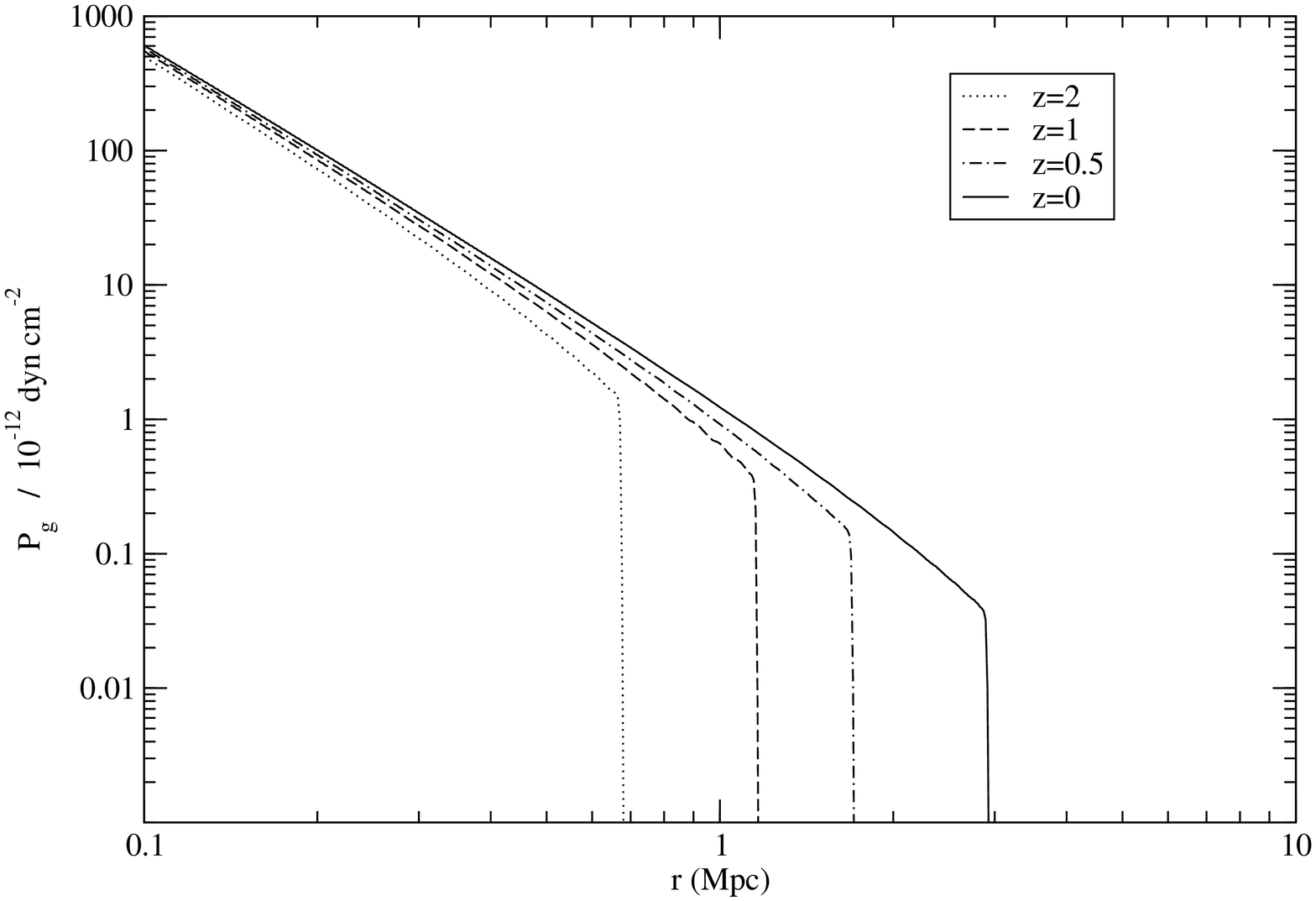}
\caption{
Gas velocity (upper panel), density (middle panel), and 
pressure (lower panel) profiles of our simulated cluster for the 
self-similar collisionless dark matter dominated accretion model in the 
Einstein--de Sitter universe at four different redshifts.  
The model has a dark matter mass 
accreted within $R_{178}$ of $M_{178} = 10^{15} M_{\odot}$ at $z=0$.
The insets in the upper and middle panels show the large radius behavior 
of the gas, with the velocity 
profile shown for $z=0$ only.  The circles on the inset of the velocity 
profile give the Hubble flow velocity at $z=0$.
We set $\Omega_b=0.05$ in this model.
} 
\label{fig:SS_sol_unscaled} 
\end{figure}

In this section, we compare our numerical simulation with the analytic 
self-similar solution of the collisionless dark matter dominated accretion 
model in the Einstein--de Sitter universe ($\Omega_M=1, 
\Omega_{\Lambda}=0$).  In this model, we assume gravity is dominated by 
the dark matter potential, and hence the self-gravity of the gas is 
switched 
off in the code.  This is done to test the ability of our code to handle 
an external gravitational potential, which is important for our realistic 
simulations with the NFW dark matter potential included.  
We set $\Omega_b$ to be 0.05 in this model.  
We assume the current Hubble constant 
to be $H_0 = 71.9$~km~s$^{-1}$~Mpc$^{-1}$ in this simulation run, and 
hence at 
redshift $z=0$, the critical density is $\rho_c(z=0) = 9.71 \times 
10^{-30}$~g~cm$^{-3}$.  The initial gas profiles are set according to the 
self-similar solution given in \citet{Ber85}.  The gravitational potential 
due to the dark matter is calculated according to the self-similar 
solution at each time step.   The gas evolution within the dark matter 
gravitational potential is then calculated by the PLUTO code.  
Figure~\ref{fig:SS_sol_unscaled} shows the dynamical variables of our 
numerical simulations for the self-similar model at four different 
redshifts.  The model shown in the three panels is such that the dark 
matter mass accreted within $R_{178}$ is $M_{178} = 10^{15} M_{\odot}$ at 
$z=0$.  The figure shows that a strong shock is propagating from $R_{\rm 
sh} \approx 0.7$~Mpc at $z=2$ to $R_{\rm sh} \approx 3$~Mpc at $z=0$.  The 
shock velocity decreases from $v_{\rm sh} \approx 1500$~km~s$^{-1}$ at 
$z=2$ down to $v_{\rm sh} \approx 1300$~km~s$^{-1}$ at $z=0$.  At $z=0$, 
the gas density just within the shock is about $0.8 \rho_c(z=0)$, and 
drops to about $0.2 \rho_c(z=0)$ just beyond the shock, while a very sharp 
jump in pressure can be seen at the shock radius.  At about $100$~Mpc, 
which is at the edge of our simulation domain, the density drops according 
to the background cosmology as $\Omega_b \rho_c$ (inset in the middle 
panel of Figure~\ref{fig:SS_sol_unscaled}).  
The velocity profile at very large radii also follows the Hubble flow 
(inset in the top panel of Figure~\ref{fig:SS_sol_unscaled}).

\begin{figure}
\includegraphics[angle=270,width=8.2cm]{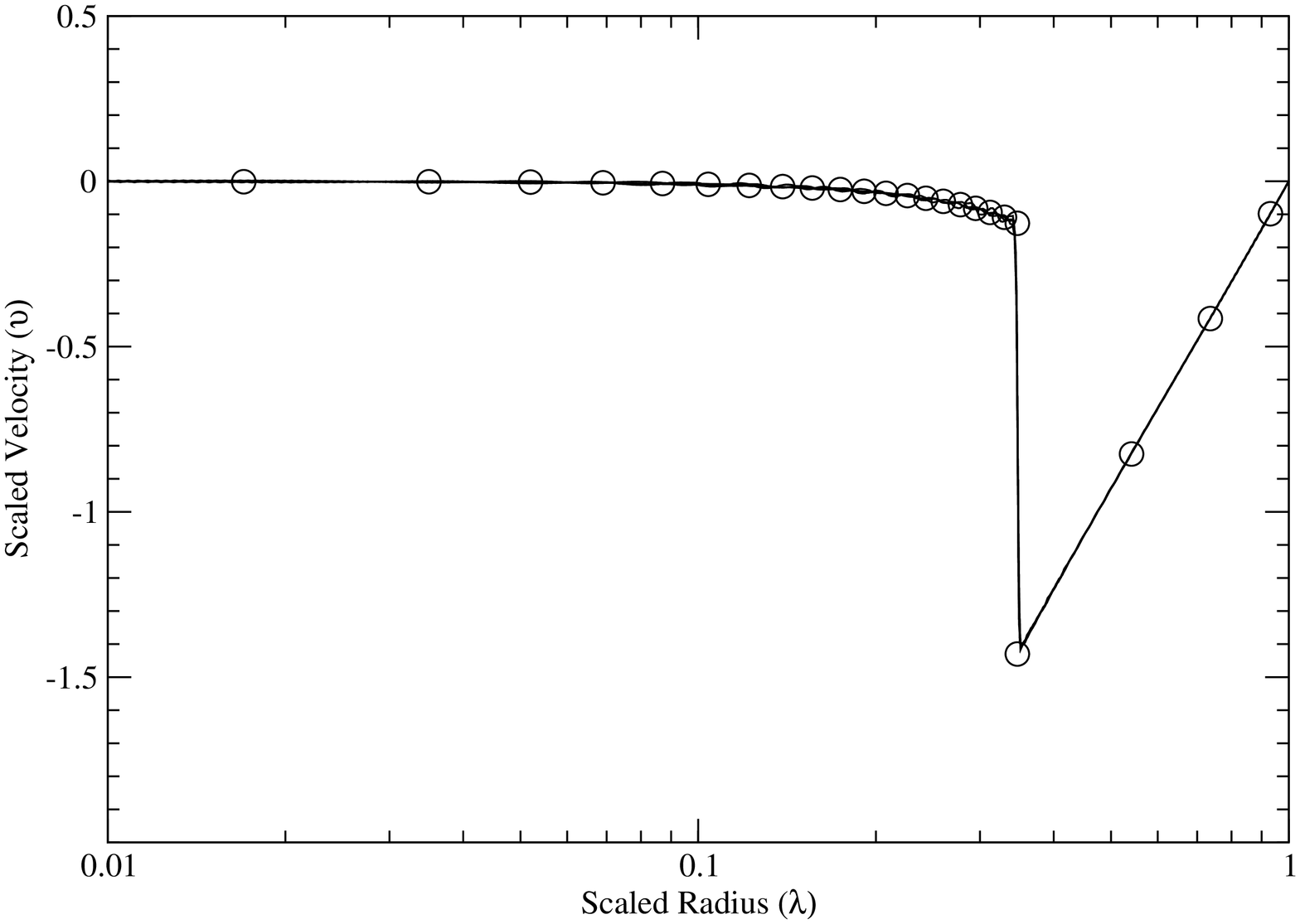}
\\
\includegraphics[angle=270,width=8.2cm]{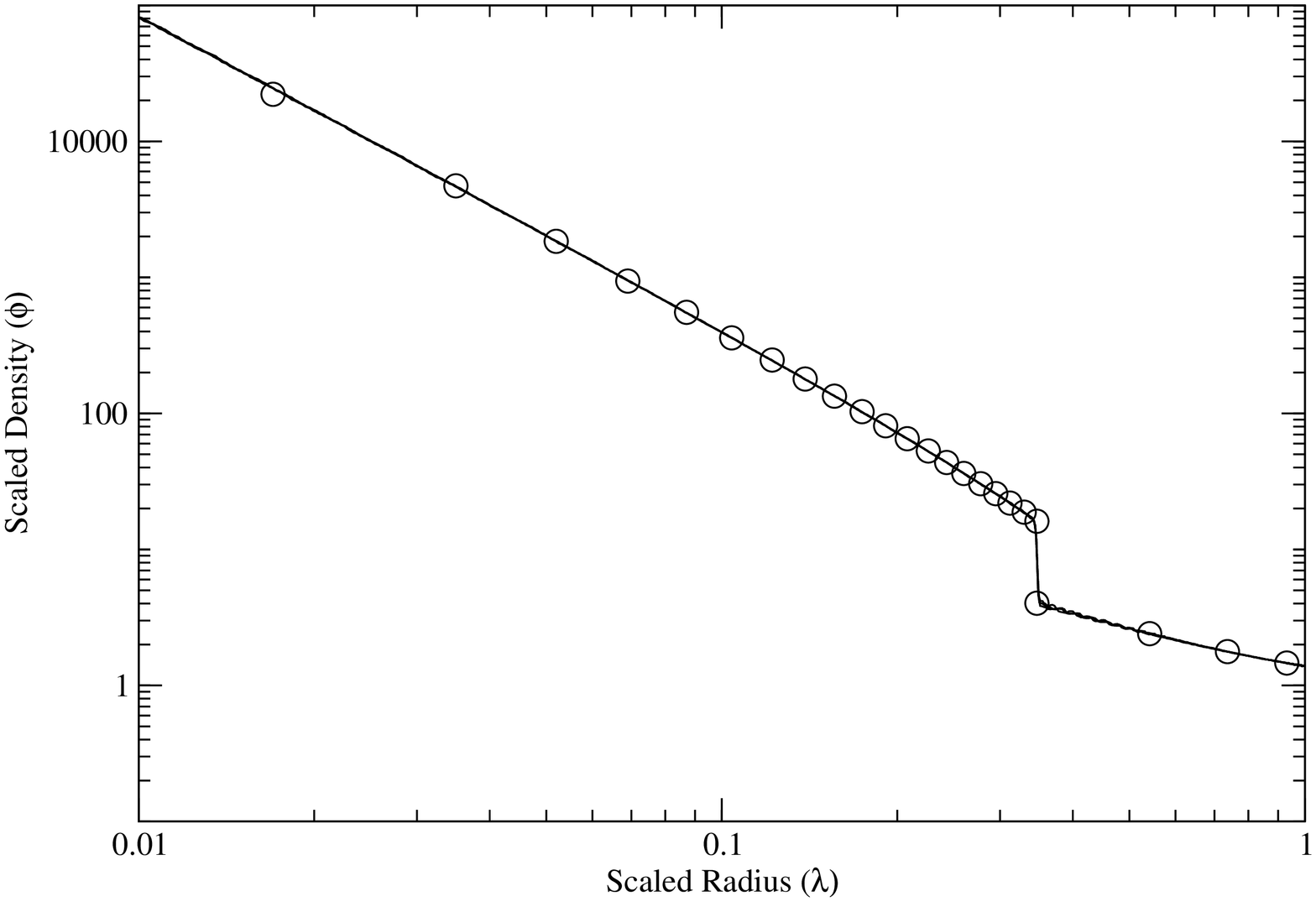}
\\
\includegraphics[angle=270,width=8.2cm]{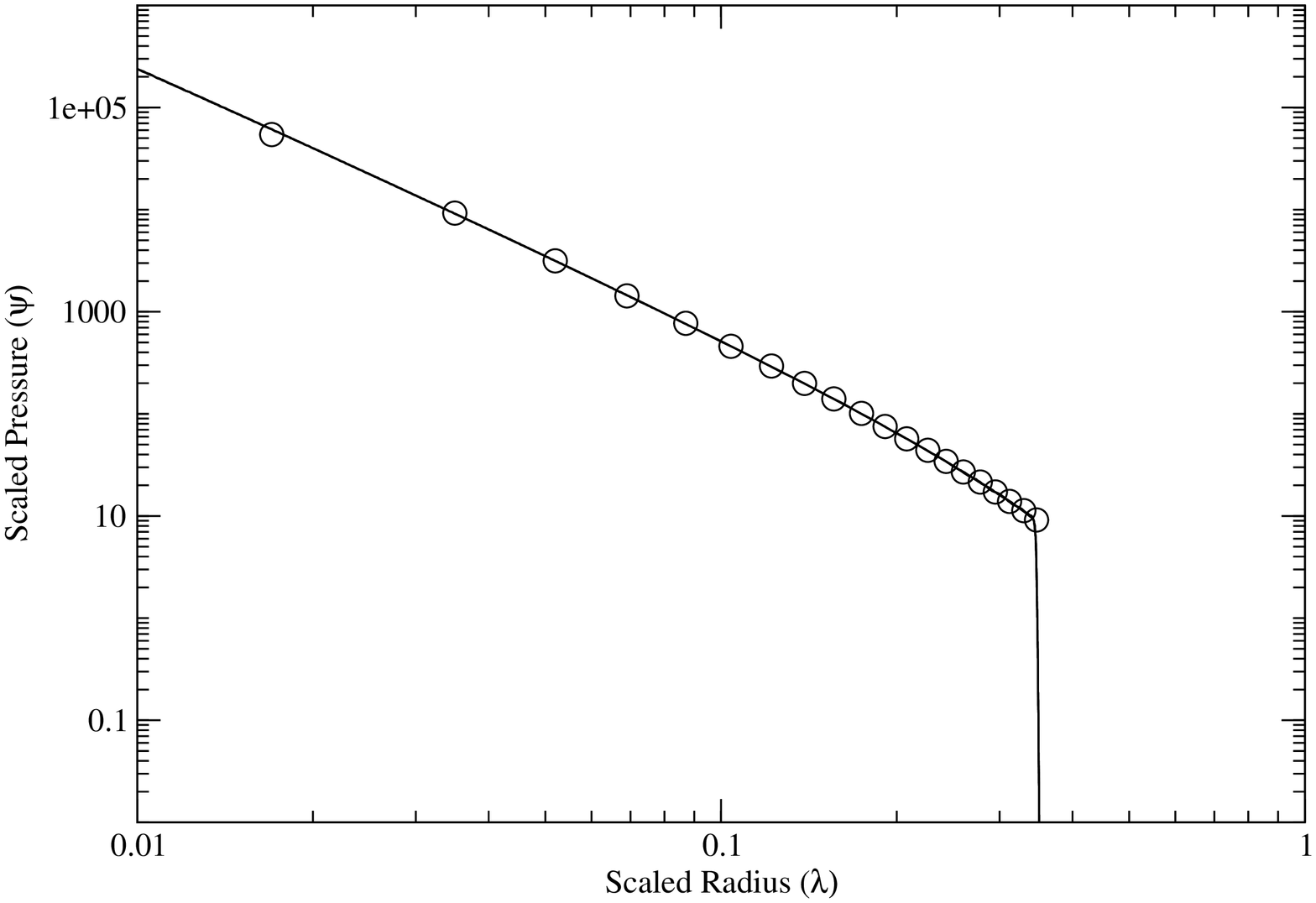}
\caption{
Scaled gas velocity (upper panel), density (middle panel), and 
pressure (lower panel) profiles of our simulated cluster.
The hydrodynamic model is the same as in Figure~\ref{fig:SS_sol_unscaled}.
All four lines lie almost on top of one another in the figures.
Circles show the corresponding analytic solutions.
} 
\label{fig:SS_sol_scaled_dm} 
\end{figure}

To compare with the analytic self-similar solution directly, we scale the 
physical radius and the dynamical variables to obtain the scaled 
dimensionless valuables according to the scaling relations given in 
\citet{Ber85},
\begin{eqnarray}
\label{eq:scaling}
\lambda & = & r/r_{\rm ta}  \, , \nonumber\\
\upsilon(\lambda) & = & \frac{v_g(r,t)}{r_{\rm ta}/t} \, , \nonumber\\
\phi(\lambda) & = & \frac{\rho_g(r,t)}{\rho_c \Omega_b} \, , \nonumber\\
\psi(\lambda) & = & \frac{P_g(r,t)}{\rho_c \Omega_b r_{\rm ta}^2/t^2} \, ,
\end{eqnarray}
where $\lambda, \upsilon, \phi$,  and $\psi $ are the dimensionless 
radius, velocity, density, and pressure of the gas, respectively, $t$ is 
the cosmic time which is equal to $1/\sqrt{6\pi G \rho_c}$ for the 
Einstein--de Sitter universe.  The scaled dimensionless valuables in our 
simulation are shown as solid lines in
Figure~\ref{fig:SS_sol_scaled_dm}.  The analytic solutions 
are plotted as open circles for comparison.  After scaling, the 
dimensionless valuables in our simulation at the four different redshifts, 
as well as the analytic solutions, all lie almost along the same locus 
which can hardly be distinguished on the figures.  This shows that our 
numerical simulations are in excellent agreement with the analytic 
solutions within all region of our interest ($\sim$~0.1--10~Mpc).  We 
conclude here that our code can calculate the dynamics of the gas, as well 
as handling the shock in an external gravity under the cosmological 
expansion correctly.

\subsection{Collisional Gas Dominated Accretion Model}
\label{sec:test_self-gravity}

\begin{figure}
\includegraphics[angle=270,width=8.2cm]{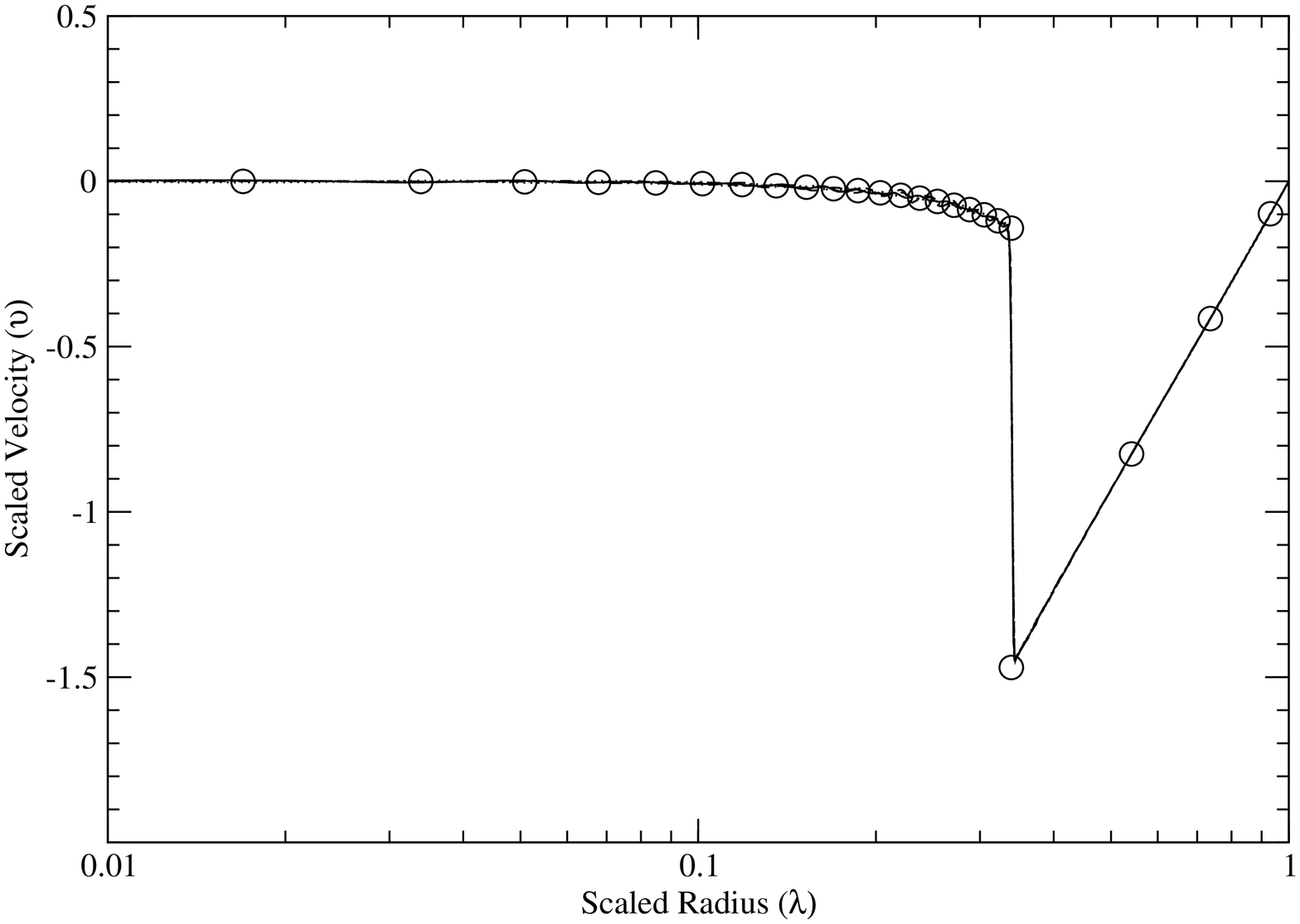}
\\
\includegraphics[angle=270,width=8.2cm]{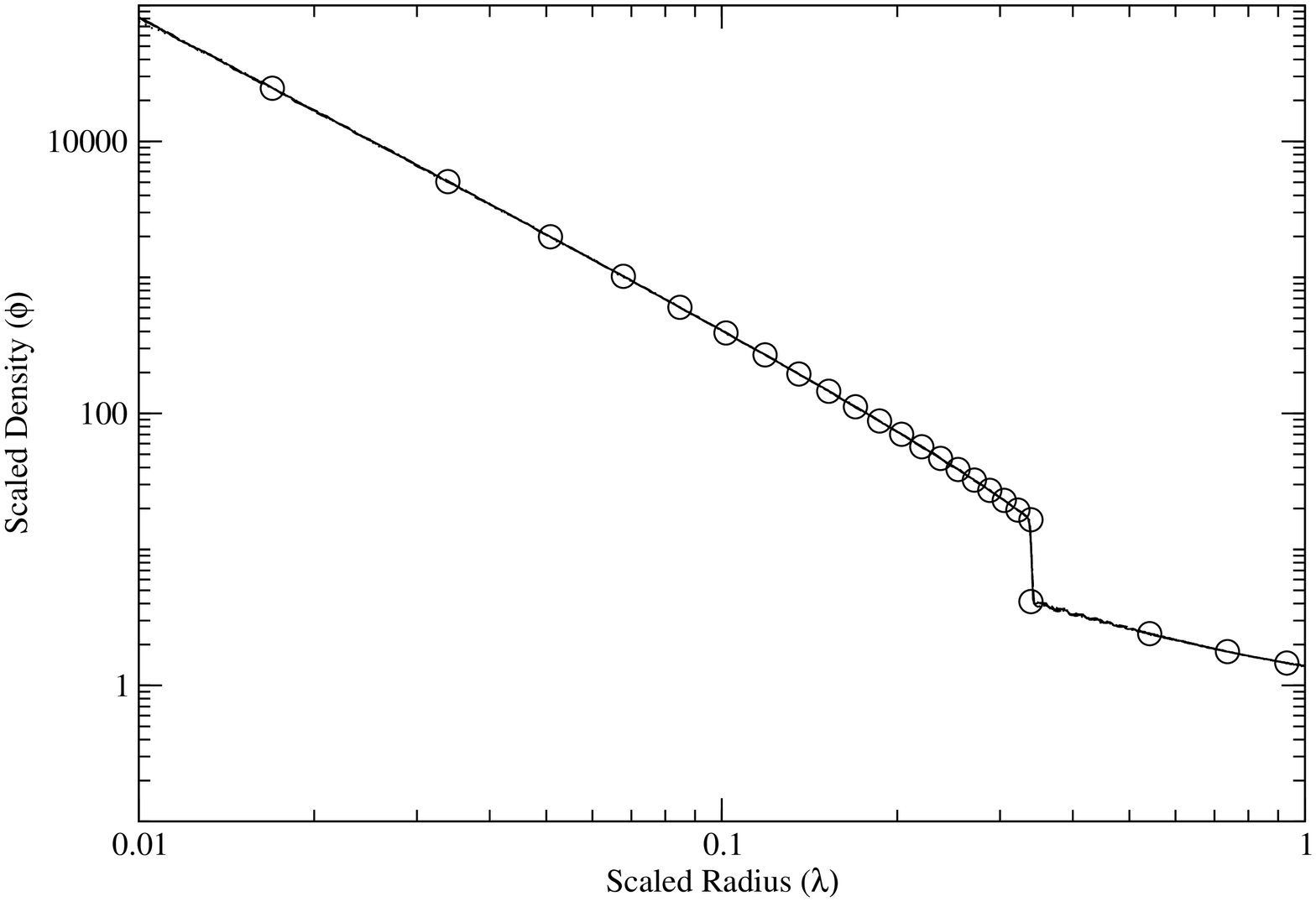}
\\
\includegraphics[angle=270,width=8.2cm]{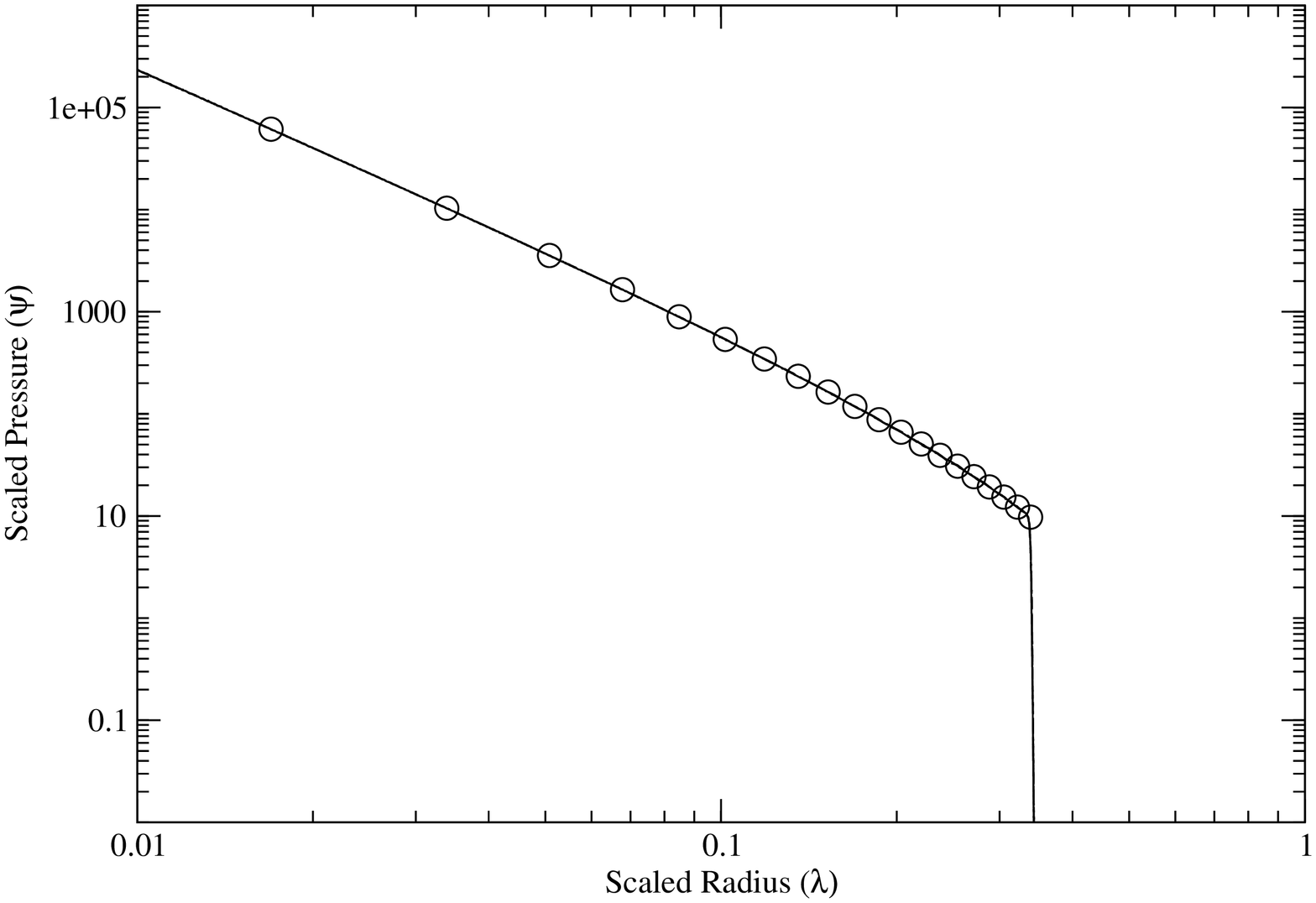}
\caption{
Similar to Figure~\ref{fig:SS_sol_scaled_dm} but for the self-similar 
collisional gas dominated accretion model in the Einstein--de Sitter 
universe.  
The model has a gas mass accreted within $R_{178}$ of $M_{178} = 1.18 
\times 10^{15} M_{\odot}$ at $z=0$.
} 
\label{fig:SS_sol_scaled_gas} 
\end{figure}

Since we have included self-gravity of the gas in our realistic 
simulations, it is necessary to check whether our code can handle 
self-gravity correctly.  We have performed a set of simulations to compare 
with the analytic self-similar solution of the collisional gas accretion 
model in the Einstein--de Sitter universe ($\Omega_M = \Omega_b=1, 
\Omega_{\Lambda}=0$).  In this model, the universe is considered as purely 
collisional fluid, and hence we do not include the dark matter 
contribution 
in these simulations.  Self-gravity of the gas is included in the 
simulations.  We scaled the physical radius and the dynamical variables 
according to equation~(\ref{eq:scaling}) with $\Omega_b = 1$.  The scaled 
dimensionless valuables in our simulations are shown as the solid lines of 
Figure~\ref{fig:SS_sol_scaled_gas}.  The analytic solutions are plotted as 
open circles for comparison.  Again, the figure shows that our numerical 
simulations are in excellent agreement with the analytic solutions.  We 
conclude here that our code can calculate the dynamics of the gas, as well 
as handle the shock with self-gravity under the cosmological expansion 
correctly.

\subsection{Self-similar Non-equipartition Model}
\label{sec:test_noneq}

\begin{figure}
\includegraphics[angle=270,width=8.2cm]{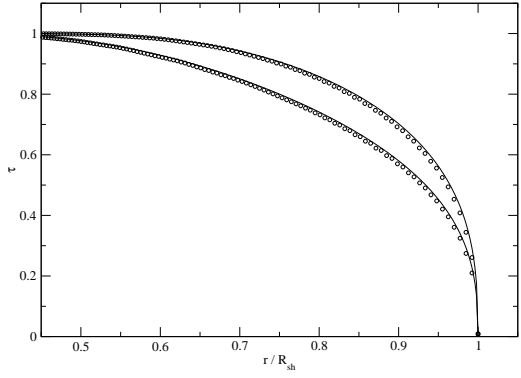}
\caption{
Ratio of the electron and average thermal dynamic temperature, 
$\tau$, as a function of scaled radius $r/R_{\rm sh}$ for clusters of the 
self-similar collisionless dark matter dominated accretion model in the 
Einstein--de Sitter universe at $z=0$.  
The upper and lower solid lines are the analytic solutions for cluster 
masses with $M_{178} = 5.26 \times 10^{14} M_\odot$ and $M_{178} = 1.05 
\times 10^{15} M_\odot$, respectively. The open circles on the 
corresponding lines are our simulated results.
We assume $\Omega_b=0.05$ here.
} 
\label{fig:test_noneq} 
\end{figure}

With the self-similar dynamical cluster model calculated in 
Section~\ref{sec:test_ex-gravity} from our simulations, we can calculate 
the degree of non-equilibration of the gas following the method given in 
Section~\ref{sec:e-heating}.  In Figure~\ref{fig:test_noneq}, the ratios 
of the 
electron and average thermodynamic temperature, $\tau$, for two clusters 
with different masses are shown as circles.
The corresponding lines are the analytic solutions 
\citep{FL97}.  Our simulations are in very good agreement with the 
analytic solutions.  The slightly deviations are mainly due to the finite 
resolution of the shock region in our numerical simulations.  We conclude 
here that our code can calculate the equilibration physics correctly.

\section{Dynamics for Realistic NFW-Dark Energy Models}
\label{sec:NFW-DE_dynamics}
\begin{figure}
\includegraphics[angle=270,width=8.2cm]{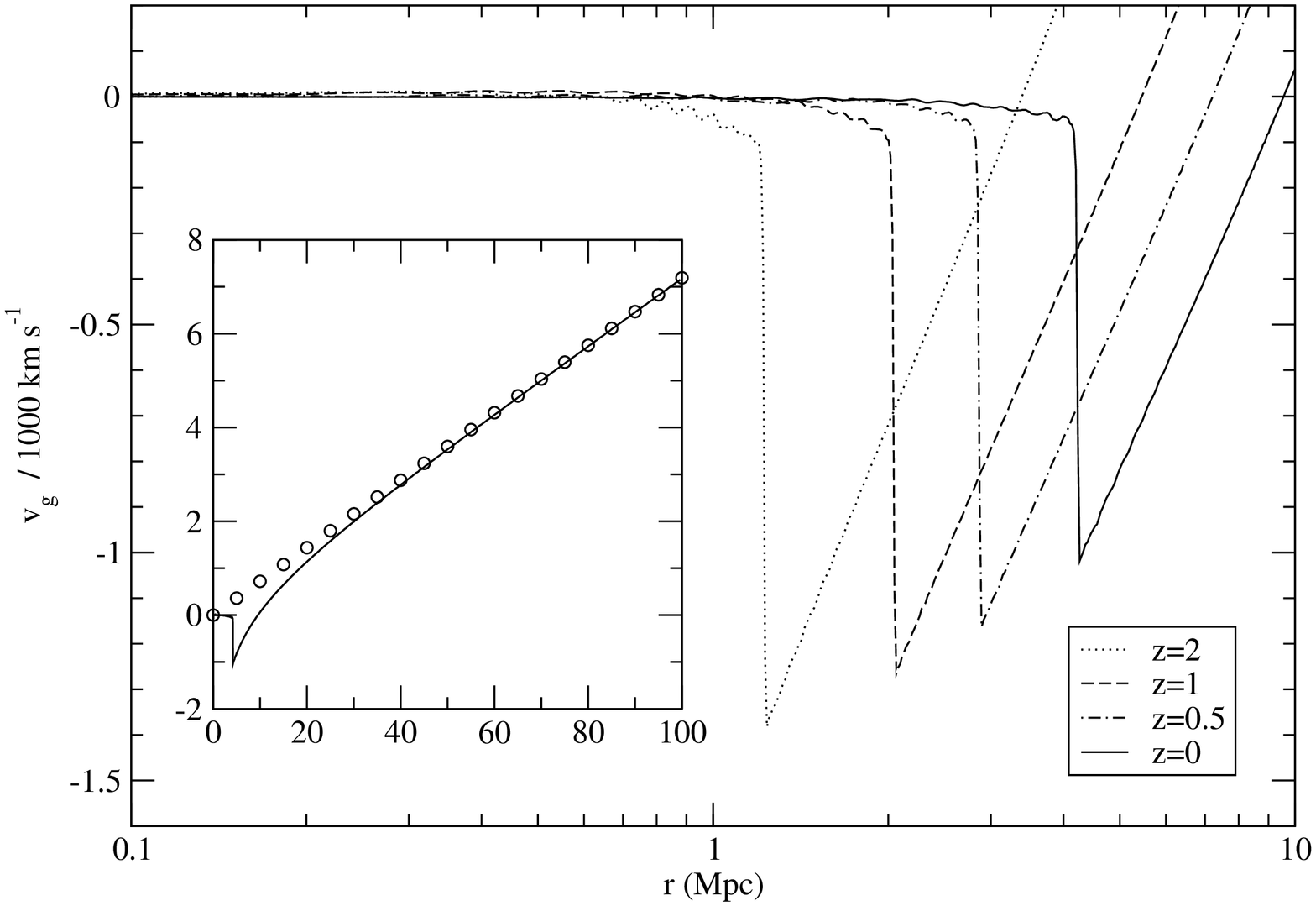}
\\
\includegraphics[angle=270,width=8.2cm]{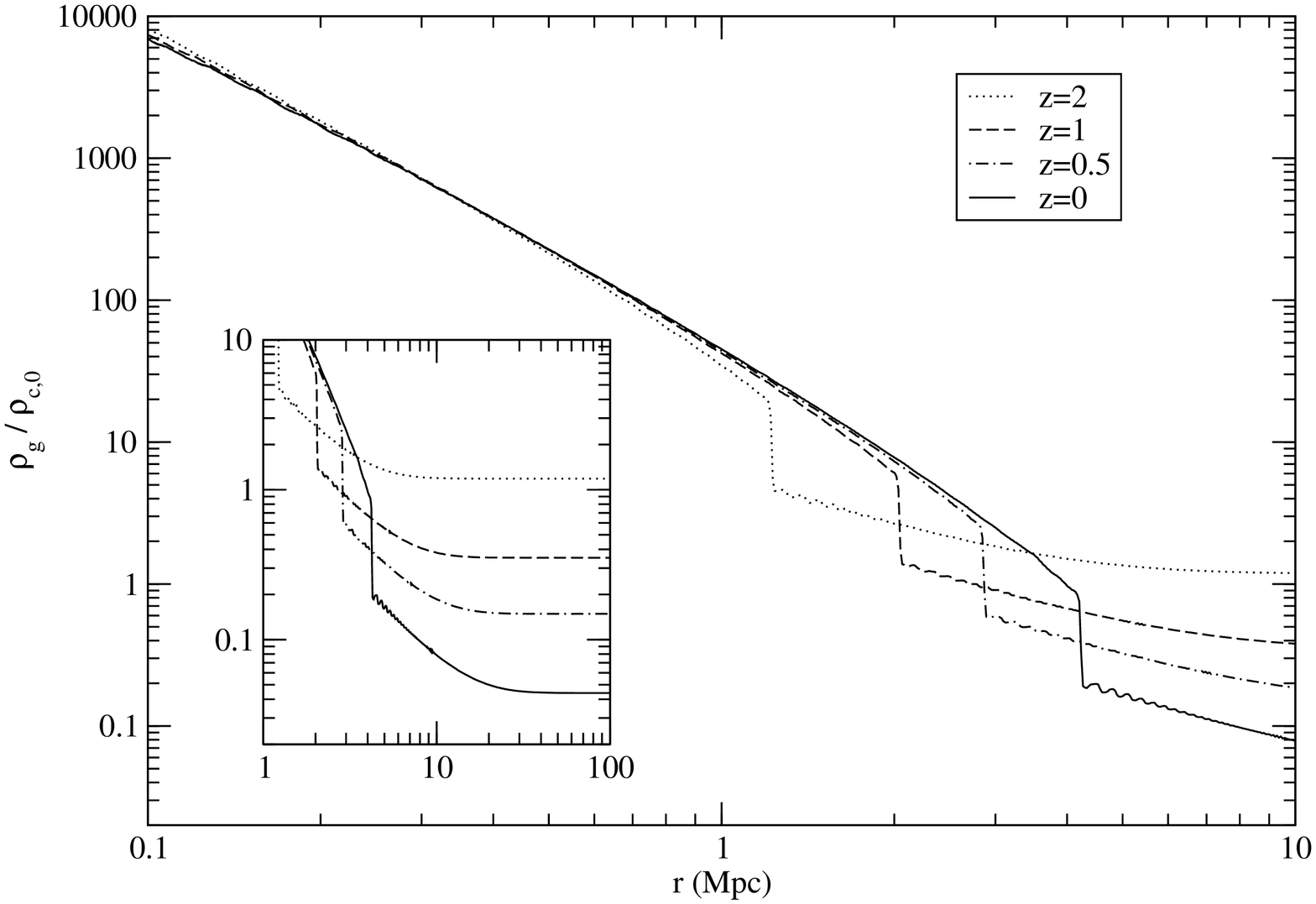}
\\
\includegraphics[angle=270,width=8.2cm]{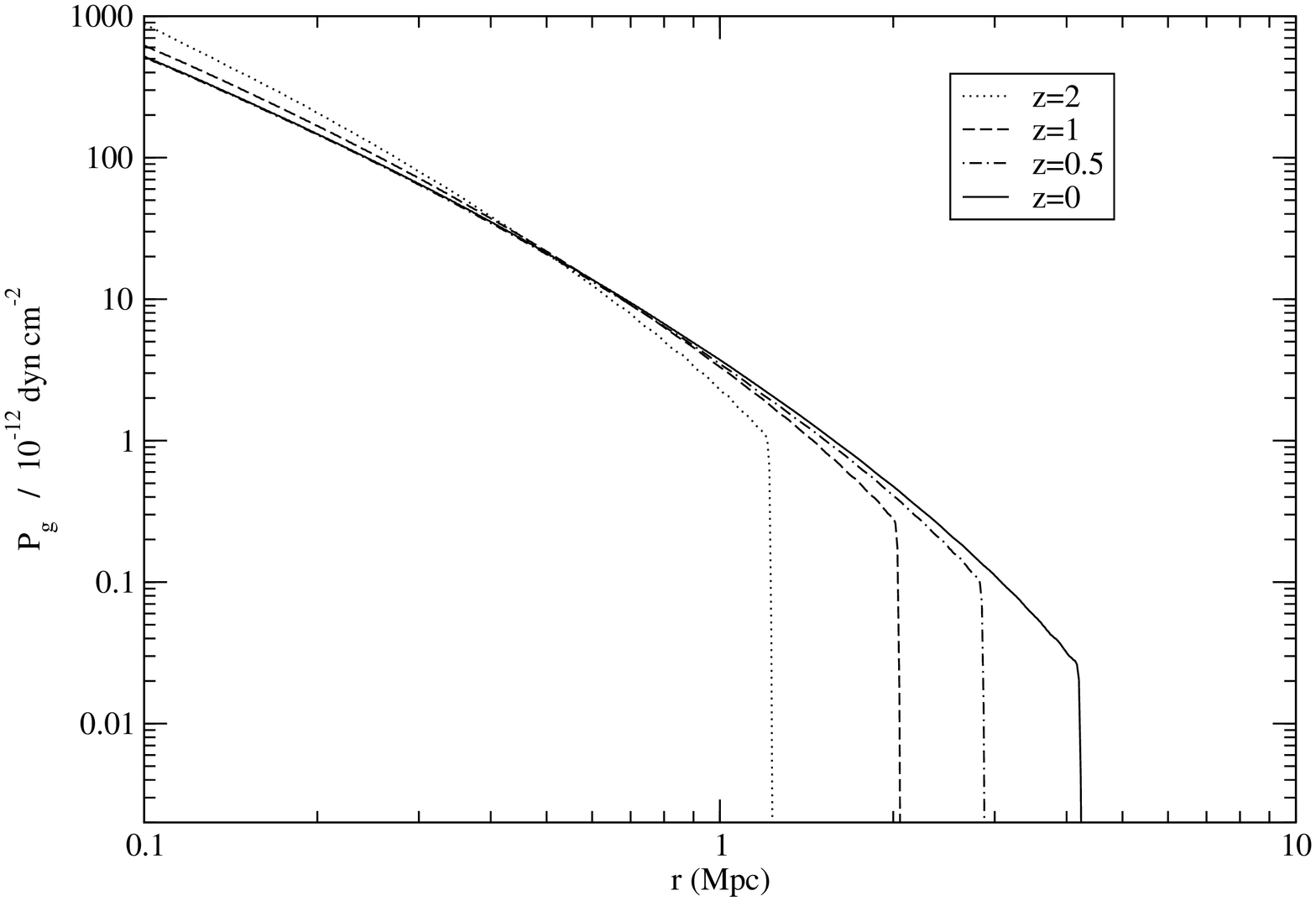}
\caption{
Gas velocity (upper panel), density (middle panel), and 
pressure (lower panel) profiles of our simulated cluster for 
the realistic NFW-dark energy model in the standard $\Lambda$CDM 
universe at four different redshifts.  
The model has total mass accreted within $R_{\rm sh}$ of $M_{\rm sh} = 
1.53 \times 10^{15} M_{\odot}$ at $z=0$.
The insets show the large radius behavior of the gas, with the velocity 
profile shown for $z=0$ only.  The circles on the inset of the velocity 
profile give the Hubble flow velocity at $z=0$.
} 
\label{fig:NFW_dyn_var} 
\end{figure}

Table~\ref{table:models} lists the masses and radii for different 
overdensities 
for some representative NFW cluster models in the standard $\Lambda$CDM 
cosmology at $z=0$.  The different definitions of masses and radii are 
used interchangeably throughout the paper. 
In Figure~\ref{fig:NFW_dyn_var}, we 
show the evolution of the dynamical variables ($v_g, \rho_g,$ and $P_g$)
as a function of radius for a cluster with an accreted mass of
$M_{\rm vir} = 
1.19 \times 10^{15} M_{\odot}$  at $z=0$, where $ M_{\rm vir} = 
M_{95}$ in the $\Lambda$CDM cosmology we assumed.  The total mass within 
the shock radius is $M_{\rm sh} = 1.53 \times 10^{15} M_{\odot}$  at 
$z=0$.  The figure shows that a strong shock propagates from $R_{\rm 
sh} \approx 1.2$~Mpc at $z=2$ to $R_{\rm sh} \approx 4.2$~Mpc at 
$z=0$.  The shock velocity decreases from $v_{\rm sh} \approx 
1400$~km~s$^{-1}$ at $z=2$ down to $v_{\rm sh} \approx 
1000$~km~s$^{-1}$ at $z=0$.  Similarly to the self-similar dark matter 
dominated model, at $z=0$, the gas density just within the shock is about 
$0.8 \rho_c(z=0)$, and drops to about $0.2 \rho_c(z=0)$ just beyond 
the shock.  At about $100$~Mpc which is at the edge of our simulation 
domain, the density drops according to the background cosmology as 
$\Omega_b \rho_c$ (inset in the middle panel of 
Figure~\ref{fig:NFW_dyn_var}).  The pressure 
profile flattens as the NFW cluster evolves, and the central pressure 
drops.  This is in contrast to the self-similar dark matter dominated 
model where the pressure never drops even in the very central region 
$< 0.1$~Mpc (not shown on the graph).  
The velocity profile at large radius also follows the 
Hubble flow correctly (the inset of Figure~\ref{fig:NFW_dyn_var}).
This shows that with our implementation of the dark energy, our 
simulations can reproduce the background cosmology correctly.
For comparison, the shock radius and the shock velocity for a self-similar 
dark matter dominated cluster with the same mass ($M_{\rm sh}$) in the 
Einstein--de Sitter universe ($\Omega_M=1, \Omega_{\Lambda}=0$) are 
$R_{\rm 
sh}^{\rm SS} = 3.11~h_{71.9}^{-2/3} (M_{\rm sh}/1.53 \times 10^{15} 
M_{\odot})^{1/3}$~Mpc and $v_{\rm sh}^{\rm SS}=1.38 \times 
10^3~h_{71.9}^{1/3}  
(M_{\rm sh}/1.53 \times 10^{15} M_{\odot})^{1/3}$ ~km~s$^{-1}$, 
respectively.
For the realistic NFW-DE model, the shock radius is larger than that of 
the self-similar solution by a factor of $1.38$, while the shock 
velocity is lower by a factor of $1.35$.  
The result is consistent with the one-dimensional $N$-body simulation 
given by \citet{RK97}.
Their results show that the shock radius and the shock velocity for a 
cluster in the $\Lambda$CDM cosmology with $\Omega_{\Lambda}\approx  0.74$ 
is about $1.4$ larger and $1.35$ lower than those in the Einstein--de 
Sitter universe, respectively.

\section{Observables for Realistic NFW-Dark Energy Models}
\label{sec:obs}
\subsection{Definition of X-ray Observables}
\label{sec:xray_obs}
X-ray spectra depend mainly on electron temperature rather than ion 
temperature, and the projected temperature profile of a cluster can be 
directly measured from X-ray observations.
Different weighting schemes have been used to calculate the projected 
temperature from hydrodynamic simulations, and \citet{MRM+04} have shown 
that most of the commonly used weighting schemes (the mass-weighted and 
the emission-weighted) give significantly different results from the X-ray 
observed spectroscopic temperature, $T_{\rm spec}$, and the 
spectroscopic-like temperature calculated from an analytic weighting 
scheme they developed is able to approximate $T_{\rm spec}$ to better than 
a few percent for temperature above $\sim 3$~keV.
However, for the 
non-equipartition model we considered, the electron temperature in the 
outer regions can be as low as $0.1$~keV.  
X-ray line emission from heavy elements contribute significantly to the 
X-ray spectrum 
for temperature below $\sim 3$~keV.  Such line emission is not 
considered by \citet{MRM+04}.  \citet{Vik06} has generalized the 
spectroscopic-like temperature to lower temperatures down to 
$\sim 0.5$~keV and to arbitrary metallicity.  This generalized scheme 
takes into account both the continuum and line emission assuming the 
MEKAL emission model \citep{MGv85,KM93, LOG95}.  The detector response has 
also been 
taken into account.  The weighting scheme is no longer analytic and has to 
be tabulated.  Following the algorithm of \citet{Vik06}, we calculate the 
projected spectroscopic-like temperature, $T_{\rm sl}$, from our numerical 
simulations.  The integration is carried out within the shock radius.  The 
latest $Chandra$ ACIS-S aim point response files released for the ACIS 
Cycle 11 proposal planning are used to generate the weighting 
table\footnote{http://cxc.harvard.edu/caldb/prop\_plan/imaging/index.html}.  
Using different $Chandra$ CCD response files does not affect the results 
significantly \citep{Vik06}.

We also calculate the surface brightness profile from our numerical 
simulations.  Here, we define $x$ as the projected radial distance from a 
cluster center.  The surface brightness profile in a given energy band $E$ 
is given by
\begin{equation}
\label{eq:SB}
S_E(x) = \int \Lambda_E(T_e,Z) n_e n_p dl
\, ,
\end{equation}
where $\Lambda_E(T_e,Z)$ is the cooling function which depends only on the 
electron temperature and heavy element abundances, $Z$, $n_p$ is the 
proton number density, and $l$ is the distance along the line of sight.  
The integration 
is carried out within the shock radius.  The cooling function 
$\Lambda_E(T_e,Z)$ is calculated using the MEKAL model \citep{MGv85,KM93, 
LOG95}.

For our models with non-equipartition considered, $T_e$ is used to 
calculate the $T_{\rm sl,non{\text -}eq}$ and $S_{E,\rm non{\text -}eq}$.  
For comparison, 
we also consider models with electrons and ions are fully in equipartition 
by taking $T_e = \bar{T}$, which is usually assumed in the literature.  
Quantities calculated with full equipartition assumed are denoted with 
the subscript ``eq''.

Note that in the outer region where electrons and ions may not be in 
equipartition, non-equilibrium ionization may also be important.  
This may increase line emissions in the soft bands ($E \lesssim 1$~keV).
In our calculations, non-equilibrium ionization is not considered.

\subsection{Definition of SZ Observables}
\label{sec:sz_obs}

The SZ effect by a cluster at $x$ can be characterized as a temperature 
increment, $\Delta T_{\rm SZE}(x)$, with respect to the CMB spectrum
\begin{equation}
\Delta T_{\rm SZE}(x) = f(\theta) y(x) T_{\rm CMB} \, ,
\end{equation}
where $\theta=h\nu / k_B T_{\rm CMB}$ is the dimensionless frequency, 
$T_{\rm CMB}$ is the CMB temperature, $y$ is the Comptonization parameter, 
and $f(\theta)$ is given by
\begin{equation}
f(\theta) = \left( \theta \, \frac{e^{\theta}+1}{e^{\theta}-1} -4 \right) 
[1+\delta_{\rm SZE}(\theta, T_e)] \, .
\end{equation}
For simplicity, we neglect the relativistic term, $\delta_{\rm SZE}$, 
which is not important for frequency lower than about 250 GHz.
The Comptonization parameter is given by
\begin{equation}
\label{eq:y-para}
y = \frac{k_B \sigma_{\rm T}}{m_e c^2} \int n_e T_e dl \propto \int P_e dl 
\, ,
\end{equation}
where $\sigma_{\rm T}$ is the Thomson scattering cross section 
and $P_e=n_e k_B T_e$ is the electron pressure.

Similarly to the calculation for the spectroscopic-like temperature, for 
our 
models in non-equipartition and in fully equipartition, we calculate the 
$\Delta T_{\rm non{\text -}eq/eq}(x)$ by integrating 
equation~(\ref{eq:y-para}) 
along the line of sight within the shock radius.

Another useful SZ observable is the integrated Comptonization parameter, 
$Y$, which is defined as the integration of the Comptonization parameter 
in equation~(\ref{eq:y-para}) on the sky
\begin{equation}
\label{eq:bigY}
 Y = d_A^2 \int y d\Omega = \int y dA,
\end{equation}
where $d_A$ is the angular diameter distance to the cluster, $\Omega$ is 
the solid angle of the cluster on the sky, and $A$ is the projected 
surface area.  We integrate the projected surface area of the cluster up 
to the shock radius.  Such a quantity is useful for spatially unresolved 
clusters with SZ observations where the solid angle covers the whole 
cluster.  

Since $Y =  \int y dA  \propto \int P_e dV \propto \int n_e T_e dV$, where 
$V$ is the volume of the cluster, the integrated Comptonization parameter 
$Y$ is basically measuring the thermal energy of the electrons in the 
cluster.  If electrons and ions are not in equipartition and if the 
electron 
temperature is lower than that of the ions globally, the value of $Y$ 
measured would be 
lower than the equipartition value.  To characterize the degree of 
non-equipartition of the whole cluster, we define the bias as the ratio 
$Y_{\rm non{\text -}eq}/Y_{\rm eq}$, where $Y_{\rm non{\text -}eq}$ is 
the integrated Comptonization parameter for the non-equilibration model 
and $Y_{\rm eq}$ is that for the equipartition model ($T_e={\bar T}$).

It has been shown that the integrated Comptonization parameter 
displays a tight correlation with cluster mass \citep{RS06}.  Such a tight 
correction is needed for precision cosmology, and 
hence a correct understanding of the integrated Comptonization parameter 
is important.  A detailed discussion of the use of SZ surveys to study 
cosmology can be found in \citet{CHR02}.

\subsection{Results for the Temperature Profiles and the X-ray 
Observables}
\label{sec:results_xray}

\begin{figure}
\includegraphics[angle=270,width=8.2cm]{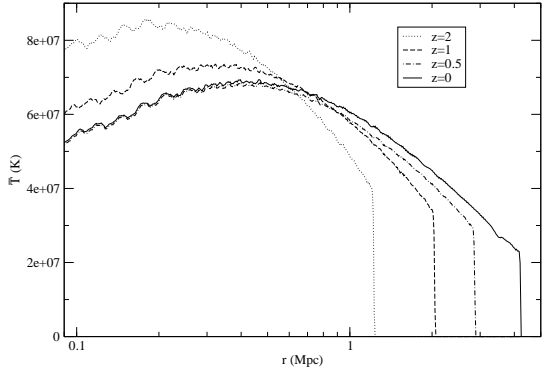}
\caption{
Average thermodynamic temperature profiles of our simulated 
realistic NFW-dark energy model cluster. 
The hydrodynamic model is the same as in Figure~\ref{fig:NFW_dyn_var}.
} \label{fig:NFW_Tbar} 
\end{figure}

\begin{figure}
\includegraphics[angle=270,width=8.2cm]{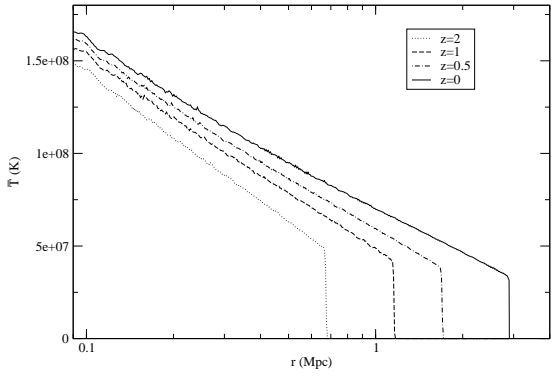}
\caption{
Average thermodynamic temperature profiles of our simulated self-similar 
dark matter dominated cluster.
The hydrodynamic model is the same as in Figure~\ref{fig:SS_sol_unscaled}.
} 
\label{fig:SS_Tbar} 
\end{figure}

The evolution of the average thermodynamic temperature as a function of 
radius for a cluster with an accreted mass of $M_{\rm sh} = 1.53 \times 
10^{15} M_{\odot}$  at $z=0$ in the $\Lambda$CDM cosmology is shown in 
Figure~\ref{fig:NFW_Tbar}.  In general, for each redshift, the temperature 
profile rises from the very central region to a peak, and then drops 
toward the outer region.  The drop of temperature in the outer region is 
due to the drop of the shock velocity during the accretion history.  The 
central drops in temperature, as well as the decrease of the value 
of the peak temperature are due to the adiabatic expansion of the cluster 
during the 
evolution.  This is in contrast to the average thermodynamic temperature 
of the self-similar solution in the Einstein--de Sitter universe where 
the temperature profile is always rising toward the central region at all 
redshift (Figure~\ref{fig:SS_Tbar}), and the temperature within the same 
gas mass, $M_{\rm gas}$, never drops during the accretion history.  
The expansion of the hot gas in the central region of the cluster in our 
simulation is probably caused by the evolution of the NFW profile, since 
there is a central pressure drop for the gas in the NFW potential but not 
for the case in the self-similar model 
(Section~\ref{sec:NFW-DE_dynamics}).
In a real cluster, the temperature profile of the central region 
($\lesssim 100$~kpc) is likely to be complicated by physical processes 
such as cooling, AGN heating \citep{Fab+00, BSM+01}, and perhaps thermal 
conduction \citep{NM01, CM04, Laz06}.
On another hand, in the outer region of a cluster, the thermodynamic state 
is likely to be dominated by gravitational processes and shock heating.  
Our models have included the essential physics in the outer regions.

\begin{figure}
\includegraphics[angle=270,width=8.2cm]{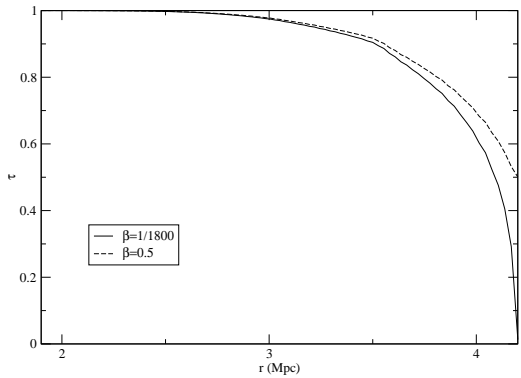}
\caption{
Ratio of electron and average thermodynamic temperatures, $\tau \equiv 
T_e/{\bar T}$, as a function of radius
at $z=0$.  
The hydrodynamic model is the same as in Figure~\ref{fig:NFW_dyn_var}.
The models with shock 
heating efficiency $\beta = 1/1800$ and $0.5$ are shown in solid and 
dashed lines, respectively.
} 
\label{fig:NFW_tau} 
\end{figure}

Figure~\ref{fig:NFW_tau} shows the ratio of electron and average 
thermodynamic temperatures, $\tau \equiv T_e/{\bar T}$, as a function of 
radius for the same cluster at $z=0$.  
The shock radius, $R_{\rm sh} \approx 4.2$~Mpc,  is about 1.5 times the 
virial radius, $R_{\rm vir} \approx 2.8$~Mpc.
Within the virial radius, the temperature differences between electrons 
and 
ions are less than $1\%$, while beyond $R_{\rm vir}$, 
$T_e/{\bar T}$ decreases from $\sim 1$ to $1/1800$ at the shock radius for 
the 
model with 
$\beta = 1/1800$.  For the $\beta = 0.5$ model, $T_e/{\bar T}$ decreases 
from $\sim 1$ to $0.5$ at the shock radius.  In general, $T_e/{\bar T}$ 
decreases from $\sim 1$ to $\beta$ at the shock radius for our models.
Our models predict that $T_e/{\bar T} \sim 0.8$ at $r \approx 0.9 
R_{\rm sh} \approx 3.8$~Mpc for a 
range of $\beta$ between $1/1800$ and $0.5$.
Beyond that radius, $T_e/{\bar T}$ depends rather strongly on $\beta$.  
The strong dependence at the shock radius can be used to distinguish shock 
heating models or constraint the shock heating efficiency of electrons at 
the shock.  

\begin{figure}
\includegraphics[angle=270,width=8.2cm]{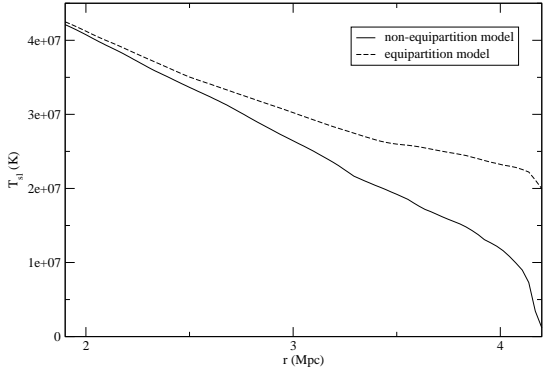}
\caption{
Projected X-ray spectroscopic-like temperature profiles 
for cluster models at $z=0$.  
The hydrodynamic model is the same as in Figure~\ref{fig:NFW_dyn_var}.
The non-equipartition model with $\beta = 1/1800$ is 
shown in solid line, while the
equipartition model is shown in dashed line.
We assume $Z = 0.3~Z_{\odot}$ here.
} 
\label{fig:NFW_Tsl_z0} 
\end{figure}

The solid line in Figure~\ref{fig:NFW_Tsl_z0} shows the projected 
spectroscopic-like temperature profiles, $T_{\rm sl,non{\text -}eq}$, of 
the 
$M_{\rm sh} = 1.53 \times 10^{15} M_{\odot}$ cluster near the outer 
region at $z=0$ for our non-equipartition model.  
We assume $\beta = 1/1800$ and $Z = 0.3~Z_{\odot}$ in this figure.
For comparison, we also 
plot the projected spectroscopic-like temperature for a model where 
equipartition of electrons and ions throughout the cluster is assumed 
($T_e = T_i = {\bar T}$) as a dashed line.  Both projected temperatures 
drop in the outer region as the radius increases, but the electron 
temperature in 
the non-equipartition model drops faster in most regions shown.  
Note that for the equipartition model, the projected temperature drop at 
the very last data point is a numerical artifact instead of a real 
feature.  This is due to the finite resolution of the shock handling in 
the hydro code which causes a slightly lower temperature  
compared to the idealized solution.  Such an artifact does not affect our 
results significantly.

\begin{figure}
\includegraphics[angle=270,width=8.2cm]{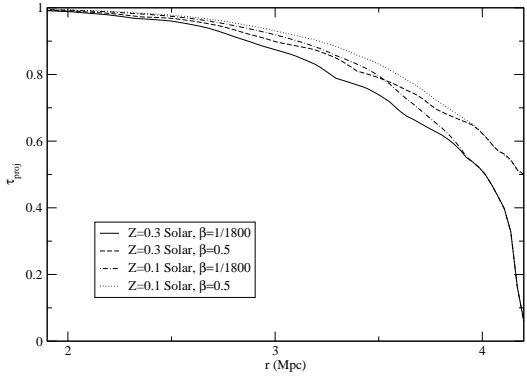}
\caption{
Ratio of the projected X-ray spectroscopic-like temperature profiles of 
the non-equipartition (with $\beta = 1/1800$ and $0.5$) and 
the equipartition models at $z=0$.  
Models with $Z = 0.1$ and $0.3~Z_{\odot}$ are presented.
The hydrodynamic model is the same as in Figure~\ref{fig:NFW_dyn_var}.
} 
\label{fig:NFW_tau_sl} 
\end{figure}

The ratio of the projected temperature profiles of 
the two models, $\tau_{\rm proj} \equiv T_{\rm sl,non{\text -}eq}/ T_{\rm 
sl,eq}$, 
is plotted in Figure~\ref{fig:NFW_tau_sl}.  
Models for $\beta = 0.5$ and $Z = 0.1~Z_{\odot}$ are also plotted for 
comparison.
Compared to the deprojected 
(or physical) temperature ratio (Figure~\ref{fig:NFW_tau}), the deviation 
is larger for the projected temperature profiles, which is  
directly determined observationally.  This is because for the projected 
temperature profile, electrons in the outer region also contribute to the 
inner region.  
For models with $\beta = 1/1800$ and $Z = 0.3~Z_{\odot}$,
there is a $\sim 10\%$ difference in the projected 
temperatures for the equipartition and the non-equipartition models at the 
virial radius ($R_{\rm vir} \approx 2.8$~Mpc) in contrast to less than a 
percent for the deprojected temperatures.  The projected temperature 
difference increases to about $20\%$ at a radius of $\sim 3.3$~Mpc, which 
is about $1.2$ of the virial radius.  
Even for $\beta = 0.5$, the difference at this radius can be as large as 
$\sim 15\%$.
We also found that the non-equipartition effect on the projected 
temperature profiles is enhanced by metallicity.  From $Z=0.1$ to 
$0.3~Z_{\odot}$, the 
deviation at a radius of $\sim 2.8 (3.3)$~Mpc is enhanced by a factor of 
$\sim 1.7 (1.5)$ for the $\beta = 1/1800$ model.  
This is because the domination of the line emissions in the soft band 
spectra is enhanced by increasing metallicity, which is more important for 
the non-equipartition model where electron temperature is lower.

\begin{figure}
\includegraphics[angle=270,width=8.2cm]{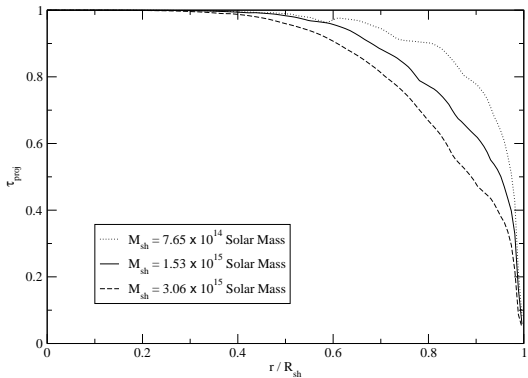}
\caption{
Ratio of the projected X-ray spectroscopic-like temperature of the 
non-equipartition with $\beta = 1/1800$ and the 
equipartition models as a function of $r/R_{\rm sh}$ 
at $z=0$ for different cluster masses.  
The hydrodynamic models are similar to those in 
Figure~\ref{fig:NFW_dyn_var}, but with $M_{\rm sh} = 7.65, 15.3, {\rm 
and~} 30.6 \times 10^{14} M_{\odot}$ at $z=0$.
We assume $Z = 0.3~Z_{\odot}$ here.
} 
\label{fig:NFW_tau_sl_diff_m} 
\end{figure}

The effect of non-equipartition is larger for more massive clusters.  
This is shown in Figure~\ref{fig:NFW_tau_sl_diff_m}, where $\tau_{\rm 
proj}$ for clusters with different masses are plotted versus $r/R_{\rm 
sh}$.  
This is because the deviation of the physical temperatures $T_e$ and 
${\bar T}$ increases with cluster mass.
The behavior qualitatively agrees with the analytic self-similar solution 
in the Einstein--de Sitter universe found by \citet{FL97}.

\begin{figure}
\includegraphics[angle=270,width=8.2cm]{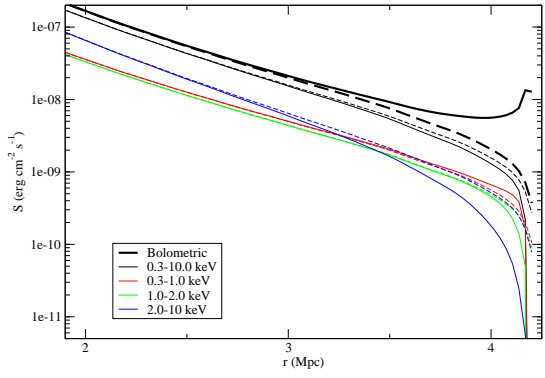}
\caption{
Surface brightness profiles for various energy bands for the 
non-equipartition (solid lines) and equipartition (dashed lines) cluster 
models at $z = 0$.
Models are the same as in Figure~\ref{fig:NFW_Tsl_z0}.
} 
\label{fig:NFW_SB}
\end{figure}

\begin{figure}
\includegraphics[angle=270,width=8.2cm]{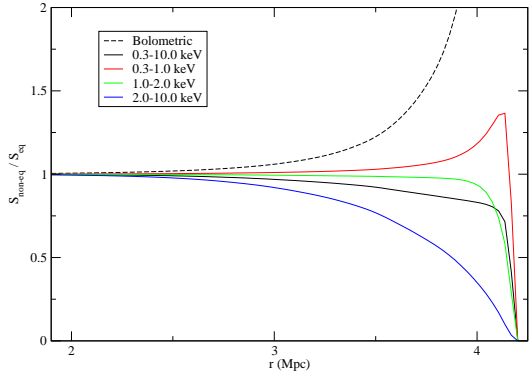}
\caption{
Ratios $S_{\rm non{\text -}eq}/S_{\rm eq}$ as a function of radius.
Models are the same as in Figure~\ref{fig:NFW_SB}.
The ratio of the bolometric surface brightness near the shock radius
reaches $\sim 35$ (outside the scale of the figure).
} 
\label{fig:NFW_SBratio}
\end{figure}

The surface brightness profiles for various energy bands of the $M_{\rm 
sh} = 1.53 \times 10^{15}M_{\odot}$ cluster near the outer region at $z = 
0$ are shown in Figure~\ref{fig:NFW_SB}.  The ratios $S_{\rm 
non{\text -}eq}/S_{\rm eq}$ are plotted in Figure~\ref{fig:NFW_SBratio}.  
We assume $\beta = 1/1800$ and $Z = 0.3~Z_{\odot}$ in these figures.  For 
X-ray emission below $\sim 2$~keV, the relative difference between the 
non-equipartition and the equipartition models is smaller than that of the 
projected temperature profile.  This is because the surface brightness 
depends on density squared but has a weaker dependence on temperature.  
The relative difference for X-ray emission above $\sim 2$~keV is similar 
to that of the projected temperature profile because of the energy cut off 
for lower temperature.  The differences in surface brightness in all 
energy bands shown are $\lesssim 10\%$ for radii $\lesssim 3$~Mpc.  Beyond 
$\sim 3$~Mpc where the electron temperature drops significantly below 
$\sim 3$~keV for the non-equipartition model, the differences in surface 
brightness profiles become important.  The difference in X-ray surface 
brightness is most significant for the hard band (2.0--10.0~keV) 
compared to the soft (0.3--1.0~keV) and medium (1.0--2.0~keV) bands.  
For the soft band, the surface brightness for the non-equipartition model 
actually becomes larger than that of the equipartition model at large 
radii.  This is because of the increase in the soft band emissivity for 
temperature below $\sim 1$~keV.  We have also plotted the full X-ray band 
(0.3--10.0~keV) and the bolometric surface brightness profiles in 
Figure~\ref{fig:NFW_SB}.  The full X-ray band surface brightness profile 
for the non-equipartition model is always lower than that of the 
equipartition model, while the opposite is true for the bolometric surface 
brightness profile.  This indicates that for the non-equipartition model, 
a large amount of emission occurs
in the energy band below 0.3~keV compared to 
that of the equipartition model.  In fact, the bolometric surface 
brightness near the shock radius for the non-equipartition model can reach 
$\sim 35$ times that of the equipartition model.

Current X-ray observations by $Suzaku$ of cluster outer regions extend to 
only $R_{200} \sim 2$~Mpc where the non-equipartition effects on both the 
surface brightness and projected temperature profiles are $\lesssim 1\%$ 
in our models.  The sensitivities of these observations are limited by the 
Poisson variations in the background extragalactic source density rather 
than by the instrumental background \citep{Bau+09}.  To push the 
sensitivity limit out to $\sim R_{\rm vir} \approx 1.4 R_{200}$ where 
non-equipartition effects on the projected temperature are $\sim 10\%$, 
observations combining high spatial resolution and high surface brightness 
sensitivity 
would be needed.  On the other hand, increasing the coverage of solid 
angle can also help to reduce the Poisson variations.  Here, we estimate 
how much improvement in sensitivity would be needed to push the current 
limit of $\sim R_{200}$ out to $R_{\rm vir} \approx 1.4 R_{200}$.  Assume 
most of the current observations only have sensitivities out to $R_{200}$ 
\citep[but see][]{GFS+09} and cluster surface brightness scales as  
shown in Figure~\ref{fig:NFW_SB}.  From Figure~\ref{fig:NFW_SB}, 
$S(R_{200})\sim 6~S (1.4 R_{200})$ for the 0.3--10.0~keV band, and 
hence to have a significant detection of X-ray emission at $R_{\rm 
vir}\approx 1.4 R_{200}$, a factor of $\sim 6$ improvement in sensitivity 
would be needed.  Probably a combination of the above two solutions (a 
factor of 6 improvement in each case) is needed to push the sensitivity 
limit to $\sim 1.4 R_{200}$.  Pushing the limit to $R_{\rm sh}$ will be 
very 
challenging with X-ray observations.  Recently, \citet{GFS+09} reported 
that cluster emission has been detected out to $\sim 1.5 R_{\rm 200}$ in 
the cluster PKS0745-191.  They have shown that the temperature at that 
radius 
is $\sim 30\%$ lower than the temperature predicted by hydrodynamic 
simulations.  If their results are confirmed, we suggest that this may be 
a signature of electron--ion non-equipartition.

\subsection{Results for the SZ Temperature Decrement}
\label{sec:results_sz1}

Compared to X-ray observations, it is believed that cluster outer regions 
should be better studied by future SZ observations because the SZ effect 
depends only on the electron density to the first power, while X-ray 
emission depends on the electron density squared.  The X-ray signature 
drops much faster compared to the SZ signature.  Moreover, the SZ effect 
is independent of redshift, and a large number of high-redshift clusters 
should be observable.  It has been suggested that future SZ observations 
such as 
the Atacama Large Millimeter Array (ALMA) 
should be able to detect or rule out the presence of accretion shocks in 
clusters 
\citep{KHF05, MHH+09}. 

\begin{figure}
\includegraphics[angle=270,width=8.2cm]{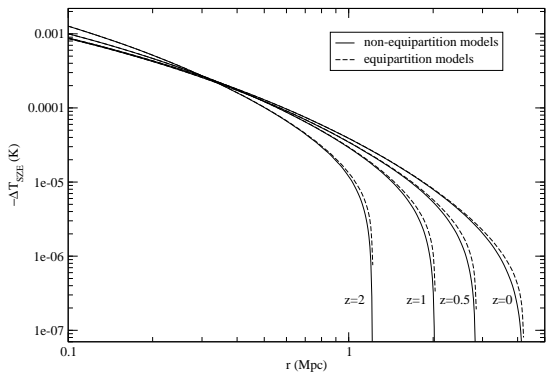}
\caption{
SZ temperature decrement profiles, $-\Delta T_{\rm SZE}$, of our simulated 
cluster
at four different redshifts.  
The non-equipartition models ($\beta = 1/1800$) are shown in solid lines, 
while the equipartition models are shown in dashed lines.
The hydrodynamic model is the same as in Figure~\ref{fig:NFW_dyn_var}.
} 
\label{fig:NFW_DeltaSZT}
\end{figure}

\begin{figure}
\includegraphics[angle=270,width=8.2cm]{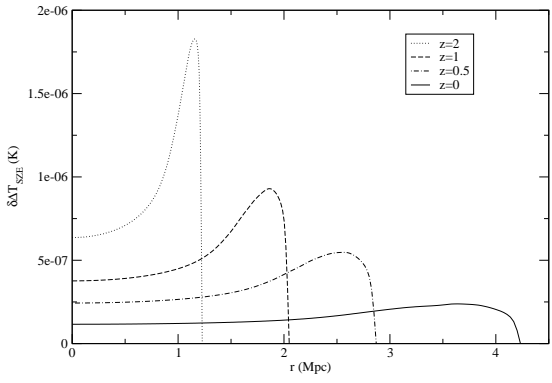}
\caption{
Difference $\delta\Delta T_{\rm SZE}$ between the SZ temperature 
decrements of the equipartition and 
the non-equipartition models ($\beta = 1/1800$)
at four different redshifts.  
Models are the same as in Figure~\ref{fig:NFW_DeltaSZT}.
} 
\label{fig:NFW_delDeltaSZT} 
\end{figure}

\begin{figure}
\includegraphics[angle=270,width=8.2cm]{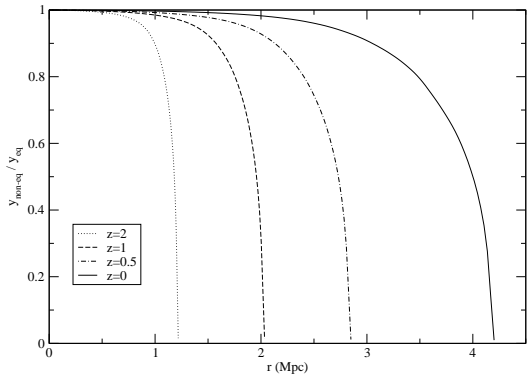}
\caption{
Ratio $y_{\rm non{\text -}eq} / y_{\rm eq}$ between the Comptonization 
parameters 
of the equipartition and the 
non-equipartition models ($\beta = 1/1800$)
at four different redshifts.  
Models are the same as in Figure~\ref{fig:NFW_DeltaSZT}.
} 
\label{fig:NFW_ypoy} 
\end{figure}

Figure~\ref{fig:NFW_DeltaSZT} shows the evolution of the 
temperature decrement magnitude due to the SZ effect, $-\Delta T_{\rm 
SZE}$, as a function of 
radius 
for a cluster with mass accreted to $M_{\rm sh} = 1.53 \times 10^{15} 
M_{\odot}$ at $z=0$.  The solid line shows the non-equipartition model and 
the dotted line shows the equipartition model.  Both the equipartition and 
non-equipartition models show very similar SZ temperature decrement 
profiles, with the SZ temperature decrement dropping faster in the 
non-equipartition models.  For the cluster at $z=0$, $-\Delta T_{\rm SZE}$ 
drops from $\sim 1$~mK at about 0.1~Mpc down to $\sim 1~\mu$K at about 
3.5~Mpc, and then drops very rapidly beyond that.  For the same cluster at 
$z=2$ when it had a mass of $M_{\rm sh}(z=2) = 6.72 \times 10^{14} 
M_{\odot}$ (corresponding to $M_{\rm vir} = M_{169} = 5.30 \times 
10^{14} M_{\odot}$ at that redshift), 
the sharp drop in $-\Delta T_{\rm SZE}$ 
occurs from a higher value of about $10~\mu$K near the shock radius.  This 
suggests that the 
shock feature can be best studied through high-redshift clusters provided 
that the region of interest can be spatially resolved.  
To show the signature of non-equipartition effect more clearly, we plot 
the difference between the SZ temperature decrements for the equipartition 
and the non-equipartition models, $\delta\Delta T_{\rm SZE} = \Delta 
T_{\rm SZE,non{\text -}eq} - \Delta T_{\rm SZE,eq}$, in 
Figure~\ref{fig:NFW_delDeltaSZT}.  It shows that the difference is larger 
at higher redshift for a given cluster.  At $z=2$, $\delta\Delta T_{\rm 
SZE}$ is of the order of $1~\mu$K for the  $M_{\rm sh}(z=2) = 6.72 
\times 10^{14} M_{\odot}$ cluster we considered.
The ratios of the temperature decrements (or equivalently the 
Comptonization parameters), $y_{\rm non{\text -}eq} / y_{\rm eq} = \Delta 
T_{\rm 
SZE,non{\text -}eq} / \Delta T_{\rm SZE,eq}$ , are also plotted in 
Figure~\ref{fig:NFW_ypoy}.  
Similarly to the temperature profiles, for $z=0$
at the virial radius of $R_{\rm vir}\approx 2.8$~Mpc, $y_{\rm 
non{\text -}eq} / 
y_{\rm eq}$ is about $0.93$, while it drops to about $0.8$ at a radius of 
$\sim 3.5$~Mpc.

\begin{figure}
\includegraphics[angle=270,width=8.2cm]{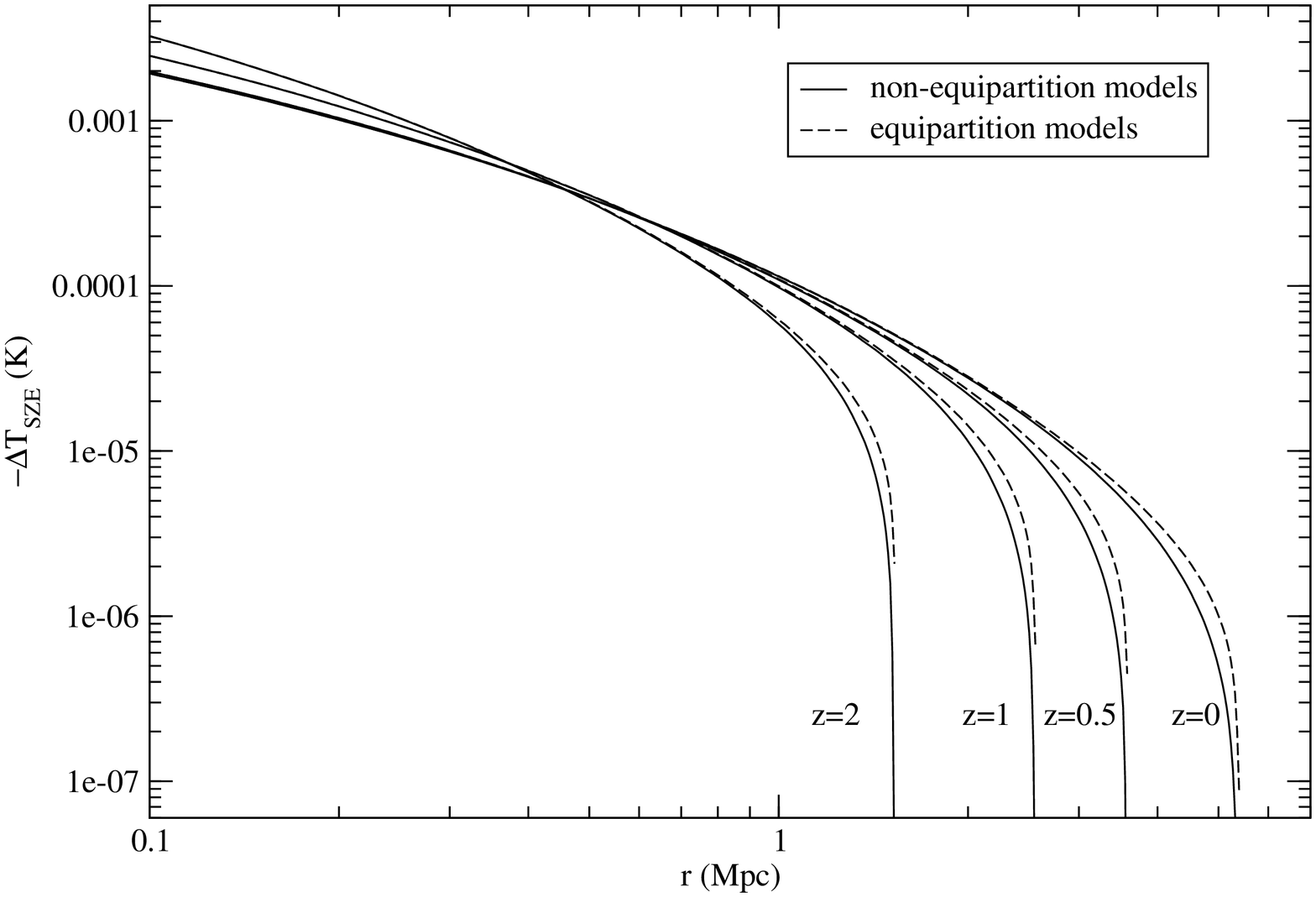}
\caption{
Same as Figure~\ref{fig:NFW_DeltaSZT} but with $M_{\rm sh}(z=0) = 
3.06 \times 10^{15} M_{\odot}$ for the cluster model.
}
\label{fig:NFW_DeltaSZT_massive}
\end{figure}

\begin{figure}
\includegraphics[angle=270,width=8.2cm]{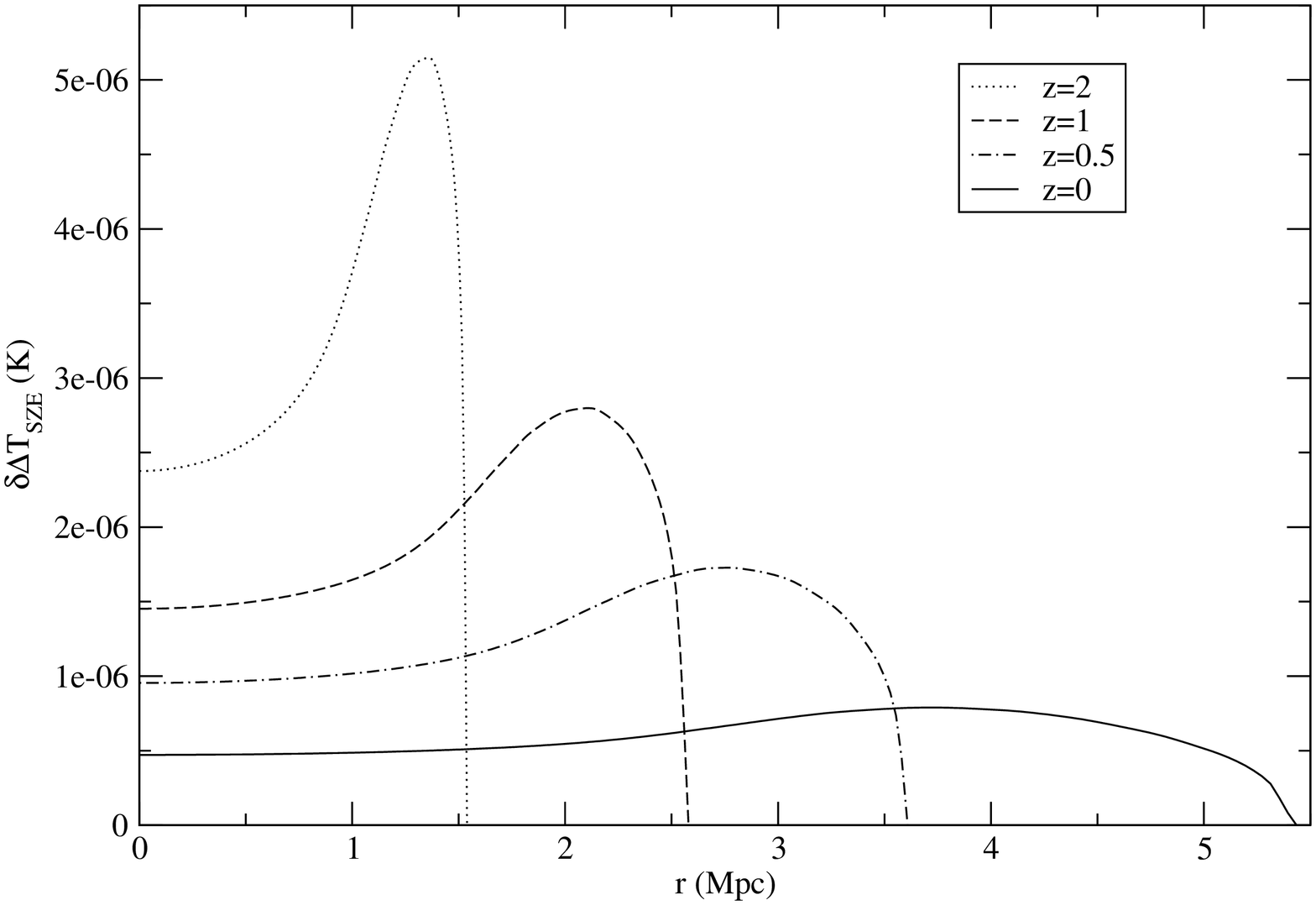}
\caption{
Same as Figure~\ref{fig:NFW_delDeltaSZT} but with $M_{\rm sh}(z=0) = 
3.06 \times 10^{15} M_{\odot}$ for the cluster model.
}
\label{fig:NFW_delDeltaSZT_massive}
\end{figure}

The detailed analysis of whether the effect can be distinguished 
observationally involves a discussion of the detail characteristic 
of the potential radio observations, which 
will be given in an upcoming paper.  Here, we only estimate roughly the 
possibility of whether the non-equipartition signature can be detected at 
the shock region.  We closely follow the estimation done by \citet{KHF05} 
and \citet{MHH+09}.  In their work, they studied whether two different 
models with and without a shock can be distinguished in future ALMA 
observations from the SZ effect.  In principle, we can adopt their 
technique directly to distinguish our two models with clusters in 
equipartition or not. In Figure~10(c) of \citet{MHH+09}, they have argued 
that for two cluster models with $\delta\Delta T_{\rm SZE}$ of the order 
of a few $\mu$K, the signal-to-noise ratio (S/N) for distinguishing 
between their models can be as high as 70.  For our model with the 
$M_{\rm sh}(z=0) = 1.53 \times 10^{15} M_{\odot}$ cluster, $\delta\Delta 
T_{\rm SZE}$ is less than $\sim 2~\mu$K even at $z=2$.  However, the most 
massive cluster today can reach a mass as high as $M_{200}(z=0) \approx 
2 \times 10^{15} M_{\odot}$.  In Figure~\ref{fig:NFW_DeltaSZT_massive}, 
we show the $-\Delta T_{\rm SZE}$ profiles for the equipartition and the 
non-equipartition models of a cluster with $M_{200}(z=0) = 1.83 \times 
10^{15} M_{\odot}$, which corresponds to $M_{\rm sh}(z=0) = 3.06 \times 
10^{15} M_{\odot}$.  The profiles are similar to those of the $M_{\rm 
sh}(z=0) = 1.53 \times 10^{15} M_{\odot}$ cluster, except the magnitudes 
are larger.  The shock radius at $z=2$ is about $1.5$~Mpc, which 
corresponds to an angular size of 177\arcsec\ for the $\Lambda$CDM 
cosmology we assumed.  The mass within the shock radius at $z=2$ is equal 
to $M_{\rm sh}(z=2) = 1.34 \times 10^{15} M_{\odot}$.  
Figure~\ref{fig:NFW_delDeltaSZT_massive} shows the deviation in $\Delta 
T_{\rm SZE}$ between the equipartition and non-equipartition models.  At 
$z=2$, the deviation is $\delta\Delta T_{\rm SZE} \approx 4-5~\mu$K 
near the shock radius.  Closely following \citet{KHF05} and 
\citet{MHH+09}, we estimate the S/N for distinguishing 
between the equipartition and the non-equipartition models of the very 
massive cluster at $z=2$ by ALMA.  
\citet{KHF05} estimated that a $\sim 20$~hr on-source integration time for 
ALMA would be enough to achieve a sensitivity of $\lesssim 10\mu$K, but 
the latest ALMA Sensitivity 
Calculator\footnote{\scriptsize 
http://www.eso.org/sci/facilities/alma/observing/tools/etc/index.html} 
shows that about $260$~hr would be needed for the required angular 
resolution (FWHM) of 2\arcsec\ at 100~GHz with 64 antennas.  
Though such a very long observation is possible, we lower the 
required angular resolution (FWHM) to 4\arcsec\ to increase 
the sensitivity, and the required on-source integration time is reduced to 
$\sim 17$~hr.
Hence we assume a 
$\sim 17$~hr on-source integration time 
and estimate the S/N ratio to be 
$\approx N_{\rm pix}^{1/2} ({\rm S/N})_1$, where $N_{\rm pix }$ is the 
number of independent pixels in the region of interest, and $({\rm 
S/N})_1$ is 
the S/N for a single pixel in ALMA.  
From Figure~\ref{fig:NFW_delDeltaSZT_massive}, the width of the annular 
region of interest from $\sim 1$ to $\sim 1.5$ Mpc is about $0.5$~Mpc 
$\sim 59\arcsec$~at $z=2$, where $1$~Mpc corresponds to $118\arcsec$ at 
this redshift for the standard $\Lambda$CDM universe assumed.  Assuming a 
Gaussian beam with an angular diameter (FWHM) of $4\arcsec$ 
for ALMA, 
the number of beams 
within the 
region of interest are about $N_{\rm pix} \approx 3000$. 
In a real 
cluster, the accretion would likely not to be spherical, and material 
would accrete through filaments.  However, in regions other than the 
filaments, accretion shocks are roughly spherical and those are the 
regions 
which our model may be applied.  \citet{MHH+09} have estimated the area 
coverage factor of such spherical accretion shock region to be $\sim 
50\%$.  With this correction, $N_{\rm pix,50\%} \approx 1500$.  The noise 
for a single pixel is estimated to be $10\mu$K \citep{KHF05,MHH+09}.  We 
take the signal for a single pixel to be $S_1 \approx\delta\Delta T_{\rm 
SZE}\approx 4\mu$K (Figure~\ref{fig:NFW_delDeltaSZT_massive}).  Thus, we 
obtain an S/N of $\sim 16$.  Even if the area coverage factor 
goes down 
to only $10\%$, the S/N can still be as high as 7.  
As noted in \citet{MHH+09}, the sensitivity of interferometers is 
reduced for large-scale smooth surface density distributions, which is 
perhaps the case for the SZ effect in the cluster outer regions.  The 
S/N in 
this analysis may be overestimated.  
However, they also argue that such 
a reduction in sensitivity can be recovered by using nonlinear 
de-convolution of data from mosaic observations \citep{HVL+02,MHH+09}.  
The above estimation assumes that the cluster model parameters (e.g., 
cluster 
mass, 
shock radius) are known in advance.  In real observations trying to 
distinguish between equipartition and non-equipartition models, if those 
parameters cannot be obtained by other means, we may need to fit the 
parameters from the data.  This will in general reduce the S/N estimated 
above. 
We defer a more detailed study to a future 
paper, but based on the 
rather high S/N estimated from the
arguments above, we conclude here that future SZ observations, such as 
those done by 
ALMA, may be able to distinguish between equipartition and 
non-equipartition models near the shock region.  

\subsection{Results for the Integrated SZ Biases, $Y_{\rm 
non{\text -}eq}/Y_{\rm 
eq}$, and its Evolution}
\label{sec:results_sz2}

\begin{figure}
\includegraphics[angle=270,width=8.2cm]{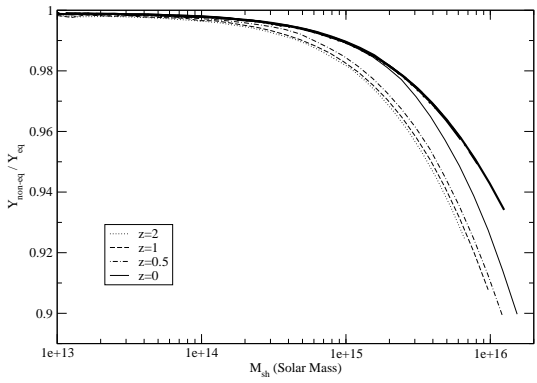}
\caption{
Integrated SZ biases, $Y_{\rm non{\text -}eq}/Y_{\rm eq}$, as a function 
of 
$M_{\rm sh}$ for both our simulated realistic NFW model in the 
$\Lambda$CDM universe (thin lines) and the numerical simulated 
self-similar model in the Einstein--de Sitter universe (thick lines) at 
four 
different redshifts.  
We assume $f_{\rm gas}=0.17$ for models in the Einstein--de Sitter 
universe.
The four lines for the self-similar model lie 
almost along the same line which cannot be easily distinguished on the 
graph.
} 
\label{fig:YBias_evol} 
\end{figure}

Figure~\ref{fig:YBias_evol} shows the integrated SZ biases, $Y_{\rm 
non{\text -}eq}/Y_{\rm eq}$ defined in Section~\ref{sec:sz_obs} as a 
function of 
$M_{\rm sh}$ for both our simulated realistic NFW model in the 
$\Lambda$CDM universe and the numerical simulated self-similar model in 
the Einstein--de Sitter universe at different redshifts.  
We assume $f_{\rm gas}=0.17$ for models in the Einstein--de Sitter 
universe 
in Figure~\ref{fig:YBias_evol}.
The masses are 
evaluated at the labeled redshift so that the evolutionary history for a 
particular cluster cannot be seen on the graph.  Both models show that 
$Y_{\rm non{\text -}eq}/Y_{\rm eq}$ increases as the cluster mass 
increases, which 
is expected as the effect of non-equipartition increases with mass.  For 
the 
realistic NFW model,  $Y_{\rm non{\text -}eq}/Y_{\rm eq}$ decreases from 
$1$ for 
$M_{\rm sh}=10^{13} M_{\odot}$ down to $\sim 0.9$ for $M_{\rm sh}=10^{16} 
M_{\odot}$ at $z=0$.  The upper mass limit to which a cluster can grow is 
limited 
by the background cosmology as well as the initial density fluctuation 
amplitudes.  The most massive nearby cluster observed has a mass of about 
$M_{200} \approx 2 \times 10^{15} M_{\odot}$, which corresponds to $M_{\rm 
sh} \approx 3.3 \times 10^{15} M_{\odot}$ in our NFW model.  $Y_{\rm 
non{\text -}eq}/Y_{\rm eq}$ for the most massive cluster in our universe 
is hence 
about $0.97$ if it is nearby ($z \approx 0$).  This bias would be larger 
($Y_{\rm non{\text -}eq}/Y_{\rm eq}$ smaller in magnitude) at higher 
redshift for a given mass.

Recent observations suggest that a significant fraction 
(20\%--40\%) of the thermal energy is missing from clusters 
\citep[][but see also \citealt{Gio+09}]{Ett03, LBC+06, VKF+06, ALN+07, 
Evr+08}.  
Obviously, if electrons and ions are in non-equipartition, the thermal 
energy 
measured by X-ray or SZ observations should be reduced.  Our 
simulations suggest that for cluster with $M_{\rm sh} \sim 1.5 
\times 10^{15} M_{\odot} (M_{200 } \sim 10^{15} M_{\odot})$, 
the non-equipartition effect can account for only about 2\%--3\% of the 
missing 
thermal energy.  For the most massive clusters, up to 3\%--4\% of the 
thermal energy beyond the equipartition value may be stored in the 
ions near the shock radius, if electrons and ions are in 
non-equipartition.  For $M_{\rm sh}$ smaller than about $5 \times 10^{14} 
M_{\odot}$ at $z=$~0--2, the non-equipartition effect is less than $1\%$.

For the self-similar model, the $Y_{\rm non{\text -}eq}/Y_{\rm eq}$ 
curves at the 
four different redshifts actually lie almost along the same line which 
cannot be easily distinguished from the graph 
(Figure~\ref{fig:YBias_evol}).  This shows that the integrated SZ bias for 
the self-similar model does not evolve with redshift.  In contrast, the 
integrated SZ bias for the realistic NFW model in the $\Lambda$CDM 
universe evolves with redshift.  In the $\Lambda$CDM universe, the 
expansion of the universe starts to accelerate around the redshift where 
$\Omega_M  \sim \Omega_{\Lambda}$; this breaks the self-similar solution 
for cosmological accretion, and hence we should expect $Y_{\rm 
non{\text -}eq}/Y_{\rm eq}$ would also deviate from self-similarity in 
general.  
For 
our realistic NFW model in the $\Lambda$CDM universe, the integrated SZ 
bias decreases as $z$ decreases.  This is probably due to the 
decreasing rate of accretion onto clusters in the $\Lambda$CDM 
universe during the cosmological acceleration, which results in a 
relatively longer timescale for the electron--ion equilibration inside a 
cluster compared to a cluster with the same mass in the 
Einstein--de Sitter universe.
At $z=0$, $Y_{\rm non{\text -}eq}/Y_{\rm eq}$ of the realistic NFW model 
is 
smaller than that of the self-similar model with the same $f_{\rm gas}$ 
for $M_{\rm sh} > 10^{15} M_{\odot}$, but the effect is similar for 
$M_{\rm sh} < 10^{15} M_{\odot}$ for both models.  At higher redshifts, 
the effect of non-equipartition is larger for the realistic NFW model for 
the entire mass range.

Though the magnitude of $Y_{\rm non{\text -}eq}/Y_{\rm eq}$ is small for 
the range 
of cluster masses, even a percentage level deviation in the most massive 
clusters is important for precision cosmology studies.  Cosmological 
studies using the mass function evolution depend sensitively on the high 
mass end of clusters.  For cluster mass count surveys using $Y$ as a 
mass proxy, if the bias in $Y$ is not properly taken into account, the 
resulted mass function would be biased low at the high mass end (i.e., 
less 
massive clusters would be observed if clusters are in non-equipartition).  
Even though the mass--$Y$ relation can be self-calibrated, the 
evolution of $Y_{\rm non{\text -}eq}/Y_{\rm eq}$ may introduce a bias if 
the 
self-calibration is not properly done at each redshift.  For example, if 
the mass and $Y$ relation is self-calibrated correctly for low-redshift 
clusters but this calibration is extrapolated to high-redshift clusters, 
bias would be introduced if the non-equipartition effect is not properly 
taken 
into account.  A detailed study of the effect of non-equipartition on 
SZ surveys and the implication to cosmological studies will be presented 
in an upcoming paper.

\section{Discussions and Conclusions}
\label{sec:conclusion}
Using one-dimensional hydrodynamic simulations, we have calculated a 
sample 
of realistic NFW clusters in a range of masses in the $\Lambda$CDM 
cosmology.  The cluster properties we simulated are consistent with the 
one-dimensional $N$-body simulations by \citet{RK97}, and they have shown 
that 
their calculations reproduce the density and temperature profiles of the 
three-dimensional simulated relaxed clusters in the outer regions.  Our 
one-dimensional hydrodynamic simulations help us to isolate the important 
physical processes under
controlled conditions.  
We have 
studied in detail the effect of non-equipartition in the outer regions of 
relaxed clusters in the $\Lambda$CDM cosmology.  

Using $f_{\rm gas} = 0.17$ (which is the upper limit for a cluster), we 
give a conservative lower limit of the non-equipartition effect on 
clusters.  We have shown that for a cluster with a mass of $M_{\rm 
sh}\sim 1.5 \times 10^{15} M_{\odot}$, within $R_{\rm vir}$, electron and 
ion temperatures only differ by less than a percent.  Our results show 
that the effect is smaller than those calculated from recent 
three-dimensional 
simulations, which shows that $T_e$ can be biased low by 5\% at $R_{200} 
\sim 0.7 R_{\rm vir}$ \citep[model CL104 in][]{RN09}.  A detailed analysis 
is needed to address the difference, but a possible explanation may be 
that 
in a three-dimensional cluster the accretion shock can be formed further 
in.  
Our results show that $T_e/{\bar T}$ can reach $\approx 0.8$ for a range 
of non-adiabatic electron heating efficiency $\beta \sim 1/1800$ to $0.5$ 
at $\sim 0.9 R_{\rm sh}$ (or $\sim 1.4 R_{\rm vir}$).  Beyond that radius, 
$T_e/{\bar T}$ depends rather strongly on $\beta$, and such a strong 
dependence at the shock radius can be used to distinguish shock heating 
models or constraint the shock heating efficiency of electron.
We also show that the effect of non-equipartition is larger for more 
massive clusters, which is consistent to analytic self-similar models in 
the Einstein--de Sitter universe \citep{FL97}.

Using the algorithm developed by \citet{Vik06} which takes into account 
the soft emission at low temperature down to $\sim 0.5$~keV, arbitrary 
metallicity, and instrumentation response, we calculated the X-ray 
spectroscopic-like temperature profiles which are the one to be directly 
determined observationally.  The effect of non-equipartition on the 
projected temperature profiles is larger than that on the deprojected (or 
physical) temperature profiles.  
Non-equipartition effects can introduce a $\sim 10\%$ bias in the 
projected 
temperature at $R_{\rm vir}$ for a wide range of $\beta$.
This is because for the projected 
temperature profile, electrons in the outer region also contribute to the 
inner region.  
We also found that the effect of non-equipartition on the projected 
temperature profiles can be enhanced by increasing metallicity.
This is because the domination of the line emissions in the soft band
spectra is enhanced by increased metallicity, which is more important for 
the non-equipartition model where the electron temperature is lower.

The effect of non-equipartition on X-ray surface brightness profile in the 
0.3--2~keV band is smaller than that on the projected temperature 
profile.  This is because the surface brightness depends on density 
squared but with a weaker dependence on temperature.  This means that in 
the outer regions, clusters in non-equipartition have similar X-ray 
surface brightness profile for $E \lesssim 2$~keV, but with bigger 
difference in temperature compared to those equipartition counterparts.
For $E \gtrsim 2$~keV, non-equipartition effects on X-ray surface 
brightness profiles are similar to those on the projected temperature 
profiles.
For a cluster with $M_{\rm sh}\sim 1.5 \times 10^{15} M_{\odot}$, 
the effect of non-equipartition on surface brightness profiles in all 
energy bands is $\lesssim 10\%$ for radii $\lesssim 3$~Mpc; beyond that, 
the effect can be important.  We found that for the non-equipartition 
model, the surface brightness profile in the low-energy band $\lesssim 
1$~keV can 
be higher than that of the equipartition model in the cluster outer 
regions.  Non-equilibrium ionization, which was not considered in our 
calculations of the emissivities, may further enhance the line emissions 
in the soft bands ($E \lesssim 1$~keV).

Current X-ray observations extend to only $\sim R_{200} \sim 2$~Mpc, 
although some results from recent $Suzaku$ observations begin to go a bit 
beyond that \citep{GFS+09}. Within those regions with $\lesssim R_{200}$, 
electrons and ions should be almost in equipartition and the signatures in 
the X-ray temperature and surface brightness should be rather weak.  But 
future X-ray observations may extend to $\sim R_{\rm vir} \approx 1.4 
R_{200}$ or even close to the 
shock radius. We have shown that non-equipartition of electrons and ions 
should be detectable in those studies.  The results by \citet{GFS+09} 
support our conclusion.

The effects of non-equipartition on the SZ effect were studied.   At 
$z=0$, 
the effect on the Comptonization parameters is similar to that of the 
projected temperature profiles.  For a cluster with $M_{\rm sh}\sim 1.5 
\times 10^{15} M_{\odot}$, $y_{\rm non{\text -}eq}/y_{\rm eq} \approx 
0.93 (0.8)$ 
at $1 (1.3) R_{\rm vir}$.  For a given cluster, the difference between the 
SZ temperature decrements for the equipartition and the non-equipartition 
models is larger at a higher redshift.  For the most massive clusters at 
$z \approx 2$,   the differences can be $\delta\Delta T_{\rm SZE} 
\approx$~4--5$~\mu$K near the shock radius.  A detailed analysis of 
whether the 
equipartition and non-equipartition models near the shock region can be 
distinguished by, for example, ALMA, will be presented in a future paper.

The effects on the integrated SZ Comptonization parameter, which measures 
the thermal energy content of the electrons, were studied. We have shown 
that the integrated SZ bias, $Y_{\rm non{\text -}eq}/Y_{\rm eq}$, 
increases as the 
cluster mass increases, which is expected as the effect of 
non-equipartition increases with mass.  
In general, the non-equipartition effect is larger for the realistic NFW 
model 
in the $\Lambda$CDM universe than that for the self-similar model in the 
Einstein--de Sitter universe, assuming that they have the same $f_{\rm 
gas}$.
Our simulations suggest that for relaxed clusters with $M_{\rm 
sh}\sim 1.5 \times 10^{15} M_{\odot}$, the non-equipartition effect can 
account for only about 2\%--3\% of the missing thermal energy globally.  
For the most massive clusters, up to 4\%--5\% of the thermal energy 
beyond the equipartition value may be stored in the thermal energy of ions 
near the shock radius, but for clusters with $M_{\rm sh} \lesssim 5 \times 
10^{14} M_{\odot}$, the non-equipartition effect is less than $1 \%$.  
Thus, 
we argue that, at least for relaxed clusters, the non-equipartition 
effect 
alone can only account for some of the missing thermal energy problem, if 
any, for high-mass clusters but not for clusters with smaller masses.  On 
the other hand, this suggests that hot gas may be missing due to other 
astrophysical processes not yet known, and the $f_{\rm gas}$ in a real 
cluster should be lower than that we used in our numerical simulations.  
We have estimated that reducing $f_{\rm gas}$ by $20\%$ will 
enhance the local non-equipartition effect near the outer region by a few 
percent, but the integrated SZ bias, $Y_{\rm non{\text -}eq}/Y_{\rm eq}$ 
is not 
affected by more than a percent.

We emphasis here that even though non-equipartition effects may not affect 
the 
global energy budget significantly, the effect is still important locally 
in the outer regions ($\sim R_{\rm vir}$) of a cluster.  Future X-ray and 
SZ 
observations may extend out to $R_{\rm sh}$, and the effect of 
non-equipartition should be considered when studying cluster properties in 
those regions.

We found that for our realistic NFW model in the $\Lambda$CDM universe, 
$Y_{\rm non{\text -}eq}/Y_{\rm eq}$ evolves with redshift, which is in 
contrast to 
the self-similar model in the Einstein--de Sitter universe.  For our 
realistic NFW model in the $\Lambda$CDM universe, $Y_{\rm 
non{\text -}eq}/Y_{\rm 
eq}$ decreases as $z$ decreases. This is probably due to the decreasing 
rate of accretion onto clusters in the $\Lambda$CDM universe 
during the period of cosmological acceleration, which results in a 
relatively longer timescale for the electron--ion equilibration inside a
cluster compared to a cluster with the same mass in the 
Einstein--de Sitter universe.
Though the 
magnitude of $Y_{\rm non{\text -}eq}/Y_{\rm eq}$ is small for the range of 
cluster 
masses, even a percentage level deviation in the most massive clusters can 
be important for precision cosmology studies.  Such a variation of $Y$ 
with $z$ would 
introduce an apparent evolution in $f_{\rm gas}$, which would bias the 
cosmological studies using the $f_{\rm gas}$ techniques \citep{ARS08}.  
Recently, \citet{RN09} have shown that the non-equipartition effect on 
$Y$ can 
be enhanced by major mergers up to $30\%$, although for low Mach number 
mergers, the shock heating efficiency for electrons may be higher 
which can weaken the non-equipartition effect \citep{GLR07, MV07}.  The 
temporary boost due to mergers may have a significant effect 
on the estimation of cosmological parameters using clusters.  
We defer a detailed study of the effect on cosmology studies in a future 
paper.

K.W. thanks Avi Loeb and Brian Mason for helpful discussions.
Support for this work was provided by the National Aeronautics and Space
Administration, through {\it Chandra} Award Numbers TM7-8010X,
GO7-8135X, GO8-9083X, and GO9-0035X, NASA $XMM-Newton$ grants
NNX08AZ34G, NNX08AW83G, and NASA $Suzaku$ grant NNX08AI27G.
We thank the referee for helpful comments.

\appendix
\section{Expression for Electron Heating within the Shock}
\label{app1}

Adiabatic changes in electron temperature are given by
\begin{equation}
\label{eq:a1}
(T_{e2})_{\rm ad} = \left( \frac{\rho_{e2}}{\rho_{e1}} 
\right)^{\gamma-1} T_{e1}
\, ,
\end{equation}
where $\rho_{e2}/\rho_{e1}=\rho_{g2}/\rho_{g1}$ for fully ionized plasma.
In addition to the adiabatic heating, electrons can also be heated by 
non-adiabatic processes.  From the definition of non-adiabatic electron 
heating efficiency (equation~(\ref{eq:beta})), the change of electron 
temperature due to non-adiabatic heating is
\begin{equation}
\label{eq:a2}
 (\Delta T_e)_{\rm non{\text -}ad} = \beta  (\Delta \bar{T})_{\rm 
non{\text -}ad}
\, .  
\end{equation}
Thus, the final electron temperature can be expressed as
\begin{equation}
\label{eq:a3}
T_{e2} = \left( \frac{\rho_{g2}}{\rho_{g1}} \right)^{\gamma-1} T_{e1} + 
(\Delta 
T_e)_{\rm non{\text -}ad}
\, .
\end{equation}
Similarly for the average thermodynamic temperature, we have
\begin{equation}
\label{eq:a4}
(\bar{T}_{2})_{\rm ad} = \left( \frac{\rho_{g2}}{\rho_{g1}} 
\right)^{\gamma-1} 
\bar{T}_{1} 
\end{equation}
and
\begin{eqnarray}
\label{eq:a5}
\left( \Delta \bar{T} \right)_{\rm non{\text -}ad} & = & \bar{T_2} - 
\left( 
\bar{T_2}\right)_{\rm ad}  \nonumber\\
& = & \bar{T_2} - \left( \frac{\rho_{g2}}{\rho_{g1}} \right)^{\gamma - 
1} 
\bar{T_1}
\, .
\end{eqnarray}
Combining equations~(\ref{eq:a2}), (\ref{eq:a3}), and (\ref{eq:a5}), we 
get
\begin{equation}
\label{eq:a6}
T_{e2} = \left(\frac{\rho_{g2}}{\rho_{g1}}\right)^{\gamma -1} T_{e1} +
\beta\,
\max\left[ 0
,\bar{T_2}  - \left(\frac{\rho_{g2}}{\rho_{g1}}\right)^{\gamma -1}
\bar{T_1} \right]
\, ,
\end{equation}
where a minimum of zero in the second term is set to ensure that numerical 
fluctuations do not introduce false decreases in entropy in the 
non-adiabatic heating.

\clearpage

\clearpage

\clearpage

\begin{deluxetable}{cccccccc}
\tabletypesize{\small}
\tablewidth{0pt}
\tablecolumns{8}
\tablecaption{ Masses and Radii$^{\rm a}$ of Some Representative NFW 
Cluster Models in the Standard $\Lambda$CDM Cosmology at $z=0$
\label{table:models}
}
\tablehead{
\colhead{$M_{\rm sh}$} & 
\colhead{$R_{\rm sh}$} & 
\colhead{$M_{\rm vir}^{\rm b}$} & 
\colhead{$R_{\rm vir}$} & 
\colhead{$M_{200}$} & 
\colhead{$R_{200}$} &
\colhead{$M_{500}$} & 
\colhead{$R_{500}$}
}
\startdata
7.65 & 3.31 & 6.05 & 2.19 & 4.90 & 1.60 & 3.59 & 1.06\\
15.3 & 4.22 & 11.9 & 2.75 & 9.50 & 1.99 & 6.83 & 1.32\\
30.6 & 5.41 & 23.1 & 3.43 & 18.3 & 2.48 & 12.9 & 1.63
\enddata
\tablenotetext{a}{  
Masses are in the unit of $10^{14}M_{\odot}$ and radii are in the unit of 
Mpc.
}
\tablenotetext{b}{  
$\Delta_{\rm vir} = 95.3$ at $z=0$ for the standard $\Lambda$CDM 
cosmology.
}
\end{deluxetable}


\begin{thebibliography}{}


\bibitem[{{Afshordi} {et~al.}(2007){Afshordi}, {Lin}, {Nagai}, \& 
{Sanderson}}]{ALN+07}
{Afshordi}, N., {Lin}, Y.-T., {Nagai}, D., \& {Sanderson}, A.~J.~R. 
2007, \mnras, 378, 293

\bibitem[{{Allen} {et~al.}(2008){Allen}, {Rapetti}, {Schmidt}, {Ebeling},
{Morris}, \& {Fabian}}]{ARS08}
{Allen}, S.~W., {Rapetti}, D.~A., {Schmidt}, R.~W., {Ebeling}, H.,
{Morris}, R.~G., \& {Fabian}, A.~C. 2008,
\mnras, 383, 879

\bibitem[{{Bautz} {et~al.}(2009)
{Bautz}, {Miller}, {Sanders}, {Arnaud}, {Mushotzky}, {Porter}, 
{Hayashida}, {Henry}, {Hughes}, {Kawaharada}, {Makashima}, {Sato}, \& 
{Tamura}}]{Bau+09}
{Bautz}, M.~W., et al. 2009, \pasj, in press (arXiv:0906.3515)


\bibitem[{{Bertschinger}(1985){Bertschinger}}]{Ber85}
{Bertschinger}, E. 1985, \apjs, 58, 39

\bibitem[{{Blanton} {et~al.}(2001){Blanton}, {Sarazin}, {McNamara}, \&
  {Wise}}]{BSM+01}
{Blanton}, E.~L., {Sarazin}, C.~L., {McNamara}, B.~R., \& {Wise}, M.~W. 
2001, \apjl, 558, L15


\bibitem[{{Borgani \& Kravtsov}(2009){Borgani}, \& 
{Kravtsov}}]{BK09}
{Borgani}, S., \& {Kravtsov}, A. 2009, Advanced Science 
Letters, in press (arXiv:0906.4370)


\bibitem[{{Bryan \& Norman}(1998){Bryan}, \& {Norman}}]{BN98}
{Bryan}, G.~L., \& {Norman}, M.~L. 1998, \apj, 495, 80
 
\bibitem[{{Bullock} {et~al.}(2001){Bullock}, {Kolatt}, {Sigad},
{Somerville}, {Kravtsov}, {Klypin}, {Primack}, \& {Dekel}}]{BKS+01}
{Bullock}, J.~S., {Kolatt}, T.~S., {Sigad}, Y., {Somerville}, R.~S.,
{Kravtsov}, A.~V., {Klypin}, A.~A., {Primack}, J.~R., \& {Dekel}, A. 2001,
\mnras, 321, 559
  
\bibitem[{{Buote} {et~al.}(2007){Buote}, {Gastaldello}, {Humphrey},
{Zappacosta}, {Bullock}, {Brighenti}, \& {Mathews}}]{BGH+07}
{Buote}, D.~A., {Gastaldello}, F., {Humphrey}, P.~J., {Zappacosta}, L.,  
{Bullock}, J.~S., {Brighenti}, F., \& {Mathews}, W.~G. 2007, \apj, 664, 
123

\bibitem[{{Bykov} {et~al.}(2008a){Bykov}, {Dolag}, \& {Durret}}]{BDD08}
{Bykov}, A.~M., {Dolag}, K., \& {Durret}, F. 2008a, Space Sci. Rev.,
134, 119

\bibitem[{{Bykov} {et~al.}(2008b){Bykov}, {Paerels}, \&
{Petrosian}}]{BPP08}
{Bykov}, A.~M., {Paerels}, F.~B.~S., \& {Petrosian}, V. 2008b, Space
Sci. Rev., 134, 141

\bibitem[{{Carlstrom} {et~al.}(2002){Carlstrom}, {Holder}, \&
{Reese}}]{CHR02}
{Carlstrom}, J.~E., {Holder}, G.~P., \& {Reese}, E.~D. 2002, \araa, 40,
643

\bibitem[{{Chandran} \& {Maron}(2004){Chandran}, \& {Maron}}]{CM04} 
{Chandran}, B.~D.~G., \& {Maron}, J.~L. 2004, \apj, 602, 170

\bibitem[{{Chieze} {et~al.}(1998){Chieze}, {Alimi}, \& {Teyssier}}]{CAT98}
{Chieze}, J.-P., {Alimi}, J.-M., \& {Teyssier}, R. 1998, \apj, 495, 630

\bibitem[{{Chuzhoy \& Loeb}(2004){Chuzhoy}, \& {Loeb}}]{CL04}
{Chuzhoy}, L., \& {Loeb}, A. 2004, \mnras, 349, L13

\bibitem[{{Dolag} {et~al.}(2002){Dolag}, {Bartelmann}, \& 
{Lesch}}]{DBL02}
{Dolag}, K., {Bartelmann}, M., \& {Lesch}, H. 2002, \aap, 387, 383

\bibitem[{{Dolag} {et~al.}(2004){Dolag}, {Bartelmann}, {Perrotta},
{Baccigalupi}, {Moscardini}, {Meneghetti}, \& {Tormen}}]{DBP+04}
{Dolag}, K., {Bartelmann}, M., {Perrotta}, F., {Baccigalupi}, C., 
{Moscardini}, L., {Meneghetti}, M., \& {Tormen}, G. 2004, \aap, 416, 853

\bibitem[{{Evrard} {et~al.}(2008){Evrard}, {Bialek}, {Busha}, 
{White}, {Habib}, {Heitmann}, {Warren}, {Rasia}, {Tormen}, {Moscardini}, 
{Power}, {Jenkins}, {Gao}, {Frenk}, {Springel}, {White}, \& {Diemand}}]
{Evr+08}
{Evrard}, A.~E., et al. 2008, \apj, 672, 122

\bibitem[{{Ettori}(2003)}]{Ett03}
{Ettori}, S. 2003, \mnras, 344. L13


\bibitem[{{Ettori \& Fabian}(1998){Ettori}, \& {Fabian}}]{EF98}
{Ettori}, S., \& {Fabian}, A.~C. 1998, \mnras, 293, L33


\bibitem[{{Fabian} {et~al.}(2000){Fabian}, {Sanders}, {Ettori}, {Taylor},
  {Allen}, {Crawford}, {Iwasawa}, {Johnstone}, \& {Ogle}}]{Fab+00}
{Fabian}, A.~C., et al.\
2000, \mnras, 318, L65

\bibitem[{{Fox \& Loeb}(1997){Fox}, \& {Loeb}}]{FL97}
{Fox}, D.~C., \& {Loeb}, A. 1997, \apj, 491, 459

\bibitem[{{Gao} {et~al.}(2008){Gao}, {Navarro}, {Cole}, {Frenk}, {White},
{Springel}, {Jenkins}, \& {Neto}}]{GNC+08}
{Gao}, L., {Navarro}, J.~F., {Cole}, S., {Frenk}, C.~S., {White},
S.~D.~M., {Springel}, V., {Jenkins}, A., \& {Neto}, A.~F. 2008, \mnras, 
387, 536


\bibitem[{{George} {et~al.}(2009){George}, {Fabian}, {Sanders}, {Young}, 
\& {Russell}}]{GFS+09}
{George}, M.~R., {Fabian}, A.~C., {Sanders}, J.~S., {Young}, A.~J., \&
{Russell}, H.~R. 2009, \mnras, 395, 657


\bibitem[{{Ghavamian} {et~al.}(2007){Ghavamian}, {Laming}, \&
{Rakowski}}]{GLR07}
{Ghavamian}, P., {Laming}, J.~M., \& {Rakowski}, C.~E. 2007, \apjl, 654,  
L69


\bibitem[{{Giodini} {et~al.}(2009){Giodini}, {Pierini}, {Finoguenov}, 
{Pratt}, {Boehringer}, {Leauthaud}, {Guzzo}, {Aussel}, {Bolzonella}, 
{Capak}, {Elvis}, {Hasinger}, {Ilbert}, {Kartaltepe}, {Koekemoer}, 
{Lilly}, {Massey}, {McCracken}, {Rhodes}, {Salvato}, {Sanders}, 
{Scoville}, {Sasaki}, {Smolcic}, {Taniguchi}, {Thompson}, \& {the COSMOS 
collaboration}}]{Gio+09}
{Giodini}, S., et al. 2009, \apj, 703, 982

  
\bibitem[{{Helfer} {et~al.}(2002){Helfer}, {Vogel}, {Lugten}, \&   
{Teuben}}]{HVL+02}
{Helfer}, T.~T., {Vogel}, S.~N., {Lugten}, J.~B., \& {Teuben}, P.~J. 2002,
\pasp, 114, 350

\bibitem[{{Hu \& Kravtsov}(2003){Hu}, \& {Kravtsov}}]{HK03}
{Hu}, W., \& {Kravtsov}, A.~V. 2003, \apj, 584, 702

\bibitem[{{Hull} {et~al.}(2001){Hull}, {Scudder}, {Larson}, \&
{Lin}}]{HSL+01}
{Hull}, A.~J., {Scudder}, J.~D., {Larson}, D.~E., \& {Lin}, R. 2001, \jgr,
106, 15711

\bibitem[{{Kaastra \& Mewe}(1993){Kaastra}, \& {Mewe}}]{KM93}
{Kaastra}, J.~S., \& {Mewe}, R. 1993, \aaps, 97, 443

\bibitem[{{Kang} {et~al.}(2007)
{Kang}, {Ryu}, {Cen}, \& {Ostriker}}]{KRC+07}
{Kang}, H., {Ryu}, D., {Cen}, R., \& {Ostriker}, J.~P. 2007, \apj, 669, 
729

\bibitem[{{Kocsis} {et~al.}(2005){Kocsis}, {Haiman}, \& {Frei}}]{KHF05}
{Kocsis}, B., {Haiman}, Z., \& {Frei}, Z. 2005, \apj, 623, 632

\bibitem[{{LaRoque} {et~al.}(2006)
{LaRoque}, {Bonamente}, {Carlstrom}, {Joy}, {Nagai}, {Reese}, \& {Dawson}, 
K.~S.}]{LBC+06}
{LaRoque}, S.~J., {Bonamente}, M., {Carlstrom}, J.~E., {Joy}, M.~K., 
{Nagai}, D., {Reese}, E.~D., \& {Dawson}, K.~S. 2006, \apj, 652, 917

\bibitem[{{Lazarian}(2006)}]{Laz06}
{Lazarian}, A. 2006, \apjl, 645, L25

\bibitem[{{Li} {et~al.}(2007){Li}, {Mo}, {van den Bosch}, \&
{Lin}}]{LMv+07}
{Li}, Y., {Mo}, H.~J., {van den Bosch}, F.~C., \& {Lin}, W.~P. 2007,
\mnras, 379, 689

\bibitem[{{Liedahl} {et~al.}(1995){Liedahl}, {Osterheld}, \& 
{Goldstein}}]{LOG95}
{Liedahl}, D.~A., {Osterheld}, A.~L., \& {Goldstein}, W.~H. 1995,
\apjl, 438, L115


\bibitem[{{Loeb}(2007)}]{Loe07}
{Loeb}, A. 2007, J. Cosmol. Astropart. Phys., JCAP03(2007)001

\bibitem[{{Makino} {et~al.}(1998){Makino}, {Sasaki}, \& {Suto}}]{MSS98}
{Makino}, N., {Sasaki}, S., \& {Suto}, Y. 1998, \apj, 497, 555

\bibitem[{{{Markevitch} \& {Vikhlinin}}(2007){Markevitch}, \& 
{Vikhlinin}}]{MV07}
{Markevitch}, M., \& {Vikhlinin}, A. 2007, \physrep, 443, 1

\bibitem[{{Mazzotta} {et~al.}(2004){Mazzotta}, {Rasia}, {Moscardini}, \&
{Tormen}}]{MRM+04}
{Mazzotta}, P., {Rasia}, E., {Moscardini}, L., \& {Tormen}, G. 2004,
\mnras, 354, 10

\bibitem[{{Medvedev}(2007){Medvedev}}]{Med07}
{Medvedev}, M.~V. 2007, \apjl, 662, L11

\bibitem[{{Mewe} {et~al.}(1985){Mewe}, {Gronenschild}, \& 
{van den Oord}}]{MGv85}
{Mewe}, R., {Gronenschild}, E.~H.~B.~M., \& {van den Oord}, G.~H.~J.
1985, \aaps, 62, 197

\bibitem[{{Mignone} {et~al.}(2007){Mignone}, {Bodo}, {Massaglia},
{Matsakos}, {Tesileanu}, {Zanni}, \& {Ferrari}}]{MBM+07}
{Mignone}, A., {Bodo}, G., {Massaglia}, S., {Matsakos}, T., {Tesileanu},
O., {Zanni}, C., \& {Ferrari}, A. 2007, \apjs, 170, 228

\bibitem[{{Molnar} {et~al.}(2009){Molnar}, {Hearn}, {Haiman}, {Bryan},
{Evrard}, \& {Lake}}]{MHH+09}
{Molnar}, S.~M., {Hearn}, N., {Haiman}, Z., {Bryan}, G., 
{Evrard}, A.~E., \& {Lake}, G. 2009, \apj, 696, 1640
  
\bibitem[{{Narayan \& Medvedev}(2001){Narayan}, \& {Medvedev}}]{NM01}
{Narayan}, R., \& {Medvedev}, M.~V. 2001, \apjl, 562, L129

\bibitem[{{Navarro} {et~al.}(1995){Navarro}, {Frenk}, \& {White}}]{NFW95}
{Navarro}, J.~F., {Frenk}, C.~S., \& {White}, S.~D.~M. 1995, \mnras, 275,
720

\bibitem[{{Neto} {et~al.}(2007){Neto}, {Gao}, {Bett}, {Cole}, {Navarro},
{Frenk}, {White}, {Springel}, \& {Jenkins}}]{Net+07}
{Neto}, A.~F., et al. 2007, \mnras, 381, 1450

\bibitem[{{Rakowski} {et~al.}(2003){Rakowski}, {Ghavamian}, \&
{Hughes}}]{RGH03}
{Rakowski}, C.~E., {Ghavamian}, P., \& {Hughes}, J.~P. 2003, \apj, 590,
846


\bibitem[{{Reid \& Spergel}(2006){Reid}, \& {Spergel}}]{RS06}
{Reid}, B.~A., \& {Spergel}, D.~N. 2006, \apj, 651, 643

\bibitem[{{Reiprich} {et~al.}(2009){Reiprich}, {Hudson}, {Zhang}, {Sato}, 
{Ishisaki}, {Hoshino}, {Ohashi}, {Ota}, \& {Fujita}}]{Rei+09}
{Reiprich}, T.~H., et al. 2009, \aap, 501, 899


\bibitem[{{Rudd \& Nagai}(2009){Rudd}, \& {Nagai}}]{RN09}
{Rudd}, D.~H., \& {Nagai}, D. 2009, \apjl, 701, L16

\bibitem[{{Ryu \& Kang}(1997){Ryu}, \& {Kang}}]{RK97}
{Ryu}, D., \& {Kang}, H. 1997, \mnras, 284, 416

\bibitem[{{Sarazin}(1986)}]{Sar86}
{Sarazin}, C.~L. 1986, Rev. Mod. Phys., 58, 1

\bibitem[{{Schekochihin} {et~al.}(2005){Schekochihin}, {Cowley}, 
{Kulsrud}, {Hammett}, \& {Sharma}}]{SCK+05}
{Schekochihin}, A.~A., {Cowley}, S.~C., {Kulsrud}, R.~M., {Hammett}, 
G.~W., \& {Sharma}, P. 2005, \apj, 629, 139

\bibitem[{{Schekochihin} {et~al.}(2008){Schekochihin}, {Cowley}, 
{Kulsrud}, {Rosin}, \& {Heinemann}}]{SCK+08}
{Schekochihin}, A.~A., {Cowley}, S.~C., {Kulsrud}, R.~M., {Rosin}, M.~S., 
\& {Heinemann}, T. 2008, {Phys. Rev. Lett.}, 100, 081301

\bibitem[{{Spitzer}(1962)}]{Spi62}
{Spitzer}, L. 1962, {Physics of Fully Ionized Gases} (2nd ed.; New York:
Interscience)

\bibitem[{{Takizawa}(1999)}]{Tak99}
{Takizawa}, M. 1999, \apj, 520, 514

\bibitem[{{Tozzi} {et~al.}(2000){Tozzi}, {Scharf}, \& {Norman}}]{TSN00}
{Tozzi}, P., {Scharf}, C., \& {Norman}, C. 2000, \apj, 542, 106

\bibitem[{{Vikhlinin}(2006)}]{Vik06}
{Vikhlinin}, A. 2006, \apj, 640, 710

\bibitem[{{Vikhlinin} {et~al.}(2006){Vikhlinin}, {Kravtsov}, {Forman}, 
{Jones}, {Markevitch}, {Murray}, \& {Van Speybroeck}}]{VKF+06}
{Vikhlinin}, A., {Kravtsov}, A., {Forman}, W., {Jones}, C., {Markevitch}, 
M., {Murray}, S.~S., \& {Van Speybroeck}, L. 2006, \apj, 640, 691

\bibitem[{{Wechsler} {et~al.}(2002){Wechsler}, {Bullock}, {Primack},
{Kravtsov}, \& {Dekel}}]{WBP+02}
{Wechsler}, R.~H., {Bullock}, J.~S., {Primack}, J.~R., {Kravtsov}, A.~V.,
\& {Dekel}, A. 2002, \apj, 568, 52

\bibitem[{{Wong} {et~al.}(2008){Wong}, {Sarazin}, {Loeb}, \& 
{Wik}}]{WSL+08}
{Wong}, K.-W., {Sarazin}, C.~L., {Loeb}, A., \& {Wik}, D.~R.
2008, in The Warm \& Hot Universe, ed F. Paerels,
http://warmhot.gsfc.nasa.gov/Posters/Poster07\_Wong.pdf

\bibitem[{{Zhao} {et~al.}(2003a){Zhao}, {Jing}, {Mo}, \&
{B{\"o}rner}}]{ZJM+03}
{Zhao}, D.~H., {Jing}, Y.~P., {Mo}, H.~J., \& {B{\"o}rner}, G. 2003a,
\apjl, 597, L9

\bibitem[{{Zhao} {et~al.}(2009){Zhao}, {Jing}, {Mo}, \&
{B{\"o}rner}}]{ZJM+09}
{Zhao}, D.~H., {Jing}, Y.~P., {Mo}, H.~J., \& {B{\"o}rner}, G. 2009,
\apj, 707, 354

\bibitem[{{Zhao} {et~al.}(2003b){Zhao}, {Mo}, {Jing}, \&
{B{\"o}rner}}]{ZMJ+03}
{Zhao}, D.~H., {Mo}, H.~J., {Jing}, Y.~P., \& {B{\"o}rner}, G. 2003b,
\mnras, 339, 12






\end{thebibliography}
\end{document}